\documentclass[preprint]{aastex631}

\newcommand{\teff}{{$T_{\rm eff}$}}
\newcommand{\logg}{{$\log g$}}
\newcommand{\vsini}{$v \sin i$}
\newcommand{\vinf}{$v_\infty$}
\newcommand{\iue}{\textit{IUE}}

\newcommand{\ha}{H$\alpha$}
\newcommand{\mdot}{{$\dot{M}$}}

\newcommand{\mdotq}{$\dot{M} q$}

\newcommand{\civ}{C~{\sc iv}}

\newcommand{\siiv}{Si~{\sc iv}}

\newcommand{\kms}{km s$^{-1}$}
\newcommand{\taurad}{{$\tau_{rad}$}}

\newcommand{\etal}{et al.}
\newcommand{\tauni}{$\tau^{(n)}_i$}
\newcommand{\rni}{$r^{(n)}_i$}
\newcommand{\wdni}{$w^{(n)}_{Di}$}

\shorttitle{Wind line variability}
\shortauthors{Massa, Prinja \& Oskinova}

\graphicspath{{./}{figures/}}

\begin{document}

\title{Wind line variability and intrinsic errors in observational mass 
loss rates}

\email{dmassa@spacescience.org}

\author[0000-0002-9139-2964]{Derck Massa}
\affiliation{Space Science Institute\\
4750 Walnut Street, Suite 205\\
Boulder, Colorado 80301, USA}

\author{Raman K. Prinja}
\affiliation{Department of Physics \& Astronomy\\
University College London \\
Gower Street, London WC1E 6BT, UK}

\author{Lidia Oskinova}
\affiliation{Institute for Physics and Astronomy\\
University of Potsdam \\
Karl-Liebknecht-Str. 24/25, D-14476 Potsdam, Germany}

\begin{abstract}
UV wind line variability in OB stars appears to be universal.  In order to 
quantify this variation and to estimate its effect on a mass loss rate 
determinated from a single observation, we use the \iue\ archive to 
identify non-peculiar OB stars with well developed but unsaturated \siiv\ 
$\lambda 1400$\ doublets and at least 10 independent observations. This 
resulted in 1699 spectra of 25 stars.  A simple model was used to 
translate the observed profile variations into optical depth variations 
and, hence, variations in measured mass loss rates.  These variations 
quantify the {\em intrinsic error} inherent in any single mass loss rate 
derived from a single observation.  The derived rates have an overall 
$\ 1 \; \sigma$\ variation of about 22\%, but this appears to differ with 
\teff, being as small at 8\% for the hottest stars and up to 45\% for the 
cooler ones.  Furthermore, any single determination can differ from the 
mean by a factor of 2 or more.  Our results also imply that mass loss 
rates determined from non-simultaneous observations (such as UV and ground 
based data) need not agree.  In addition, we use our results to examine 
the nature of the structures responsible for the variability. 
Our findings suggest that the optical depth variations result from optically 
very thick structures occulting more or less of the line of sight to the 
stellar disk. Further, the smaller optical depth variations in the hotest 
stars suggests that the structures responsible for the variations are 
disrupted in their more powerful winds.  

\end{abstract}



\section{Introduction} \label{sec:intro}

It is well known that the winds of massive stars help to power the 
interstellar medium and contribute to determining how OB stars will end 
their nuclear burning lifetime and the nature of their remnants.  
Consequently, it is important to understand the mechanisms responsible 
for mass loss and our ability to measure it.  One complication to 
achieving this is that the winds of massive stars are variable. 

\begin{figure*}
\begin{center}
\includegraphics[width=1.0\linewidth]{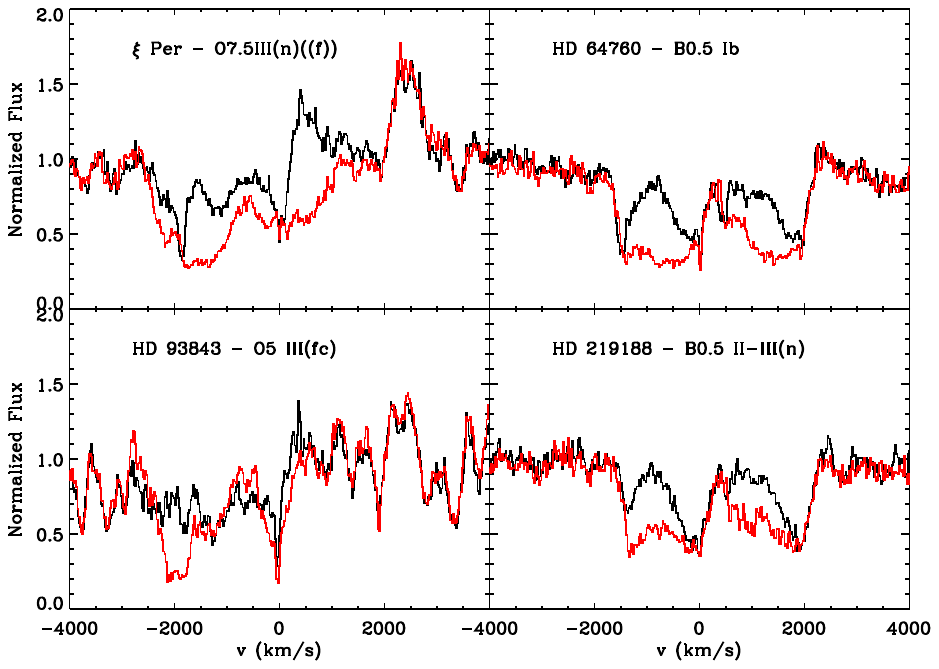}
\end{center}
\vspace{-4.3in}
\caption{Examples of \siiv $\lambda\lambda 1400$\ variability in four 
stars with a range in spectral types, demonstrating the ubiquity of wind 
line variation.}
\label{fig:var}
\end{figure*}

Wind line variability, as seen in \ha, has been known for many years 
\citep[e.g.][]{ebbets82}.  With the advent of the {\it International 
Ultraviolet Explorer}, \iue, it became clear that virtually all well 
developed but unsaturated UV wind lines in OB stars were highly variable 
\citep[e.g.][Figure~\ref{fig:var}]{prinja86, massa95, kaper96, prinja02}.  
Later studies demonstrated that the wind lines of LMC and SMC OB stars 
vary as well \citep{massa00, lehner03}, and that the wind lines of the 
central stars of planetary nebulae show similar variations 
\citep{prinja12}.  Consequently, it is important to understand the nature 
of the variability because {\em wind line variability appears to be a 
universal property of radiatively driven winds}.  

In addition to universality, studies have revealed much about the nature 
of wind line variability and its origin.  First, it is repetitive, on a 
time scale on the order of the stellar rotation period \citep{prinja88}.  
Second, once a feature is detected at low velocity, it persists to the 
terminal velocity \citep{kaper96, massa95, dejong01, prinja02}, 
suggesting that the structures are very large and probably due to spiral 
features in the wind \citep{prinjahowarth88}.  Third, the structures 
originate at or very near the stellar surface \citep{howarth95, massa15}.  
Fourth, for B supergiants is has been shown that the structures are 
optically very thick and cover a large portion of the face of the star 
\citep[e.g.][]{prinja10}.  Fifth, X-ray observations imply that the 
structures are considerably denser than the rest of the wind 
\citep{massa19}.  Sixth, when a star is viewed nearly equator on, the 
wavelength dependence of the variability presents a pattern indicative of 
large spiral structures in the wind \citep{cranmer96, fullerton97, dejong01}.  

All of these factors affect the wind diagnostics and how we convert them 
into mass loss rates.  One consequence of this is that the four different 
means to measure mass loss rates (\mdot): wind lines, IR/Radio excesses, 
X-ray emissions and bow shocks, do not all agree   
\citep[e.g.][]{fullerton06, massa17, kobulnicky19}. The most likely 
reason is that current models do not adequately represent the density 
structures (clumping) in the winds. 

Observations of wind line variability present an opportunity to 
characterize the wind structures responsible for the variability and to  
determine the intrinsic accuracy of a single measurement of \mdot.
In this paper, we address the effects of wind line variability 
quantitatively.  We consider its effect on mass loss rate diagnostics 
and what it might reveal about the physical nature and geometry of the 
structures responsible for the variability.  This is done by fitting 
repeated \iue\ observations of the same star using the Sobolev 
approximation with exact integration model, \citep[SEI,][]{lamers87} and 
then examining the variations in the optical depths of the wind lines, 
which are proportional to \mdotq, where $q$ is the ionization fraction of 
the ion being analyzed.  We also allow the ratio of $f$ values for the 
doublet to be variable, as this has been shown to be a coarse measure of 
the amount of optically thick clumping in the wind \citep{prinja10}.  

In the following, we explain why we concentrate on the \siiv\ doublet and 
present the sample of stars analyzed.  Next, we detail how the wind lines 
are fit.  We then analyze our results and discuss their implications.  A 
final section summarizes our findings.

\section{The Sample} \label{sec:sample}

To produce our sample, we began with all O and B III-I stars observed by 
\iue\ at high resolution with the short wavelength prime (SWP) camera, 
which covered the wavelength range $1150 \lesssim \lambda \lesssim 2000$
\AA\ with a resolution of $\simeq 15$\ \kms.  We did this using the \iue\ 
archive at the Mikulski Archive for Space Telescopes (MAST).  This sample 
was then restricted to stars that had 10 or more well exposed spectra.
Stars were observed repeatedly by \iue\ for a multitude of reasons.  It 
was common practice to obtain multiple exposures to increase the relatively 
low signal-to-noise of a single \iue\ spectrum.  Some stars were monitored 
for calibration purposes, and others to search for variability.  

We then eliminated all peculiar stars such as, Be stars, extremely rapid 
rotators, interacting binaries and X-ray binaries.  Finally, we limited 
the sample to stars with well developed but unsaturated \siiv\ $\lambda 
1400$ doublets.  We concentrate on this doublet because it has the largest 
doublet separation of the wind lines available to \iue.  Letting 
$\lambda_B$ and $\lambda_R$\ denote the blue and red wavelengths of a 
doublet, the velocity separation of the doublet is given by the parameter 
$\delta \equiv c (\lambda_R -\lambda_B)/ \lambda_B$ \kms.  For \siiv, 
$\lambda_B$\ and $\lambda_R$\ are 1393.755 and 1402.750 \AA, so $\delta = 
1936$\ \kms.  In contrast, it is only 499 \kms\ for the \civ $\lambda 
1550$\ doublet.  As a result of the large doublet separation, the 
components of the \siiv\ lines are less entangled \citep[see the 
discussion of common point surfaces by][]{olson}, and that will be 
important for the fitting described in \S~\ref{sec:sei}.  The restriction 
of the sample to stars with well developed but unsaturated \siiv\ wind 
lines results in a sample consisting primarily of O and early B stars 
above the main sequence \citep[see][]{walborn85, walborn95}.  

Biases in the sample are inevitable.  Many of the stars have several time 
series, spanning different time scales.  If these series span a time scale 
that is much shorter than the time scale of the intrinsic variations, then 
they will underestimate the variability since they are effectively just 
repeated observations of the star in the same state.  In addition, one must 
keep in mind that while \siiv\ may be the dominant ion in B supergiant 
winds, it is only a trace ion in O star winds, so it may be capturing a 
completely different aspect of the wind in each case.  

Table~\ref{tab:stars} presents some properties of the program stars.  It 
lists the star name, the spectral type, the number of \iue\ spectra 
available, the number of days spanned by the data, the \vsini\ of the star 
from \cite{howarth97} and the maximum stellar rotation period, $P_{max}$.  
Stellar radii are needed to calculate $P_{max}$.  The $M_V$ were 
determined using $V$, $E(B-V)$, the distance (from either {\em GAIA} or 
{\em Hipparcos}) and the assumption that $R(V) = 3.1$. Then, using the 
\teff~s derived in \S~\ref{sec:sei}, and the bolometric corrections from  
\cite{searle08} for the B stars or \cite{martins05} for the O stars, we 
obtained the stellar radii.  These are then combined with the observed 
\vsini\ to arrive at the $P_{max}$ values given in the table.  

For HD 47240, the {\em GAIA}\ data yielded an unrealistic result, implying 
an $M_V = -8.2$\ mag, and $R/R_\odot = 66.0$.  However, this is a normal 
B1 Ib whose \vsini\ of 100 \kms\ is somewhat high for a supergaint.  
Consequently, we assigned it an $M_V = -5$\ mag, which is typical for its 
spectral type, and use that to calculate $P_{max}$.   

All of the \iue\ data presented in this article were obtained from the 
Mikulski Archive for Space Telescopes (MAST) at the Space Telescope Science 
Institute. The specific observations analyzed can be accessed via 
\dataset[DOI: 10.17909/d72e-de92]{https://doi.org/10.17909/d72e-de92}.

\section{Modeling the wind line variability}\label{sec:sei}

In order to translate flux variations into the physical parameters needed 
to estimate the intrinsic error in a single \mdot\ measurement, a model is 
required.  We use the SEI Sobolev model as formulated by \cite{lamers87} 
and modified by \cite{massa03}. 
We note that the results described in the following sections imply that 
some of the assumptions inherent in the SEI model (such as spherical 
symmetry) are incorrect.  However, this is unavoidable since these same 
results suggest that no current model is strictly correct.  Another 
simplification of the SEI model is that it is a "core-halo" model, making 
it unreliable at low velocity.  However, more sophisticated models, e.g.,  
CMFGEN \citep{hillier90} or PoWR \citep{hamann98}, require introducing 
additional, ill-defined parameters and assumptions, such as micro- and 
macro- clumping \citep{oskinova06}, the velocity law of the 
wind-photospere interface and the assumption that the photosphere is 
homogeneous).  Further, these models entail the computation of millions of 
lines in NLTE, making them ill-suited for the least squares process that 
we use, since it requires many tens or more re-computations for each fit.  
Consequently, it seems reasonable to concentrate on the nature of the wind 
beyond about 1.5 stellar radii using the simplest possible model in order 
to minimize the number of free parameters.  This approach is simple and 
its output well-defined, making it a useful guide for the further 
development of the more sophisticated models.  And importantly, the SEI 
model can be calculated quickly, making it ideal for use in an automated 
fitting procedure for a large number of spectra.  

The calculation of an SEI profile requires the following parameters: a 
value for the terminal velocity, \vinf; a velocity law, typically a 
$\beta$\ law of the form $w = (1 - a/x)^\beta$, where $x = r/R_\star$, 
$w = v/$\vinf, and $a = 1-w(x=1)^{1/\beta}$; the turbulent velocity of 
the wind, $w_D$, and the radial (Sobolev) optical depth of the wind 
\begin{equation}
  \tau_{rad}(w) = Const \frac{\dot{M}}{R_\star v_{\infty}^2} 
                q_i(w) \left(x^2 w \frac{dw}{dx}\right)^{-1} 
\end{equation}
where $Const$\ contains atomic parameters and $q_i$ is the ionization 
fraction, which is assumed to be constant.  The function $\tau_{rad}(w)$ 
is modeled by 20 velocity bins adjusted to obtain the best fitting profile.  
We also set $w(x=1) = 0.01$\ throughout.  \textit{Note that $\tau_{rad}$\ 
variations are proportional to derived $\dot{M}$\ variations.}

In addition to the usual parameters, we also allow the ratio of the 
oscillator strengths of the doublets, $f_B/f_R$, to vary.  This is a well 
known means to mimic the effect of optically thick structures partially 
covering the stellar disk \citep{prinja10}.  It also provides an additional 
diagnostic of how the portion of the wind structures in front of the star 
varies with time.  We note that letting $f_B/f_R$\ be free has little 
impact on \taurad.  Its main effect is to improve the SEI fit to both 
components of the wind doublet.  All of these parameters are determined by 
a non-linear least squares fit to the observed profile.

The fitting process involved the following three steps.

\noindent {\bf 1.  Preparation of the model photospheric spectrum:} We 
used a simple visual comparison between the photoshperic lines in 
the SWP spectrum and those of a grid of TLUSTY model atmospheres described 
by \cite{hubney24}
\footnote{
The new model atmospheres have  739,791 points which cover $200$\AA$ \leq 
\lambda \leq 32 \mu$\ with an $R = 100,000$.  They include models with $15 
\leq T_{eff} \leq 55$ kK, $0.1 \leq Z/Z_\odot \leq 3.0$ and microturbulent 
velocities, $v_t$, of 2, 5 and 10 \kms.}. 
The values of \vsini\ and the radial velocity, 
$v_{rad}$, are also adjusted in this process.  We are not concerned with 
the accuracy of the atmospheric parameters (we simply require a match to 
the photospheric spectrum near the wind line), but the results are 
typically close to expectations.  The normalization of the model atmosphere 
is performed using a version of the \cite{fitz90} extinction parameters, 
and a model hydrogen atom to fit the interstellar Ly~$\alpha$\ absorption.  

\noindent {\bf 2. A grid search for \boldmath{\vinf\ and $\beta$:}} 
Using the mean spectrum and a few individual spectra we perform a grid 
search to determine the \vinf\ and $\beta$\ that produce the best fits.  
We also examine the blue edge of \civ\ $\lambda 1548$\ (which is heavily 
saturated in the program stars) as a guide.  This grid search involves 
a non-linear least squares fit using the routine developed by 
\cite{mark09} to find the best \taurad\ and $w_D$, with \vinf\ and 
$\beta$\ fixed for each fit.  For the $\beta$\ search, we placed 
additional weight on the emission portion of the profile, since it is 
formed by the entire wind and should be most indicative of the global 
properties of the wind.  We fix the photospheric model, $\beta$\ and 
\vinf\ for the star in subsequent fitting.  These parameters are 
listed in Table~\ref{tab:params} which gives the name of the star, its 
spectral type, and the \teff, \logg\ and turbulent velocity, $v_t$, of 
the model atmosphere used for the photospheric spectrum.  It also gives 
the values adopted for the radial velocity and terminal wind velocity.  
      
\noindent {\bf 3. A least squares fit for \boldmath{$\tau_{rad}(w)$}, 
$w_{D}$,\ and $f_B/f_R$:}  We then performed a non-linear least squares 
fit to each spectrum with the following independent variables: 20 
independent bins to characterize $\tau_{rad}(w)$, $w_D$\ and $f_B/f_R$.  
In the fitting, we give the emission portion of the profile 30\% of the 
weight allotted to the absorption because the emission originates 
throughout the wind, and is only weakly variable.  In contrast, we are 
interested in the highly variable absorption, which is formed in a 
cylinder between the observer and the stellar disk and is not expected 
to correspond to the global emission at any given time.

\begin{figure*}
\includegraphics[width=.6\linewidth]{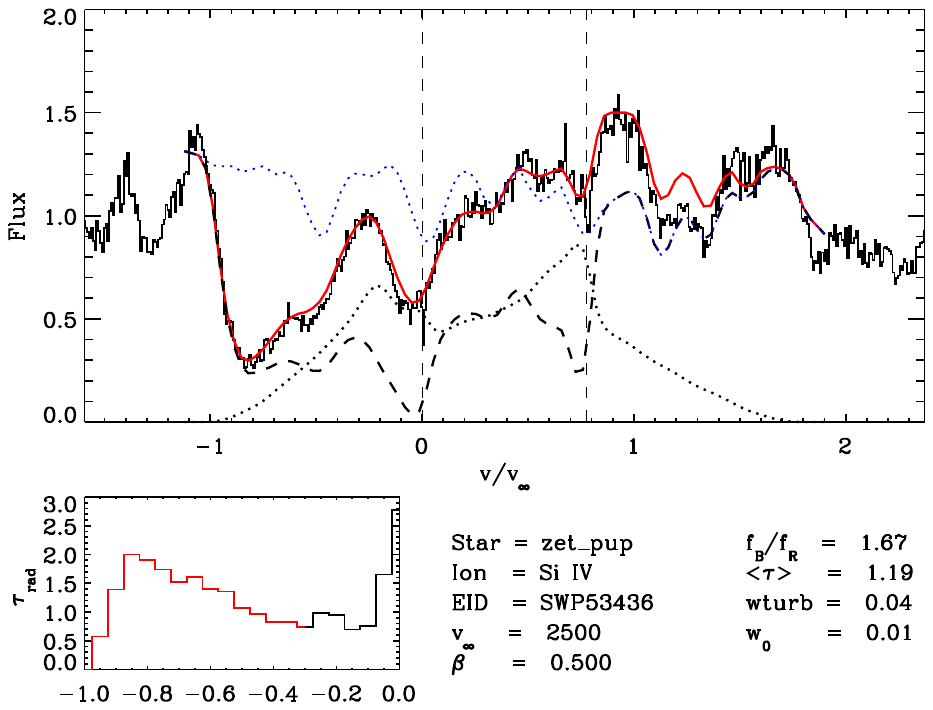}\hspace{-0.7in}\includegraphics[
width=.6\linewidth]{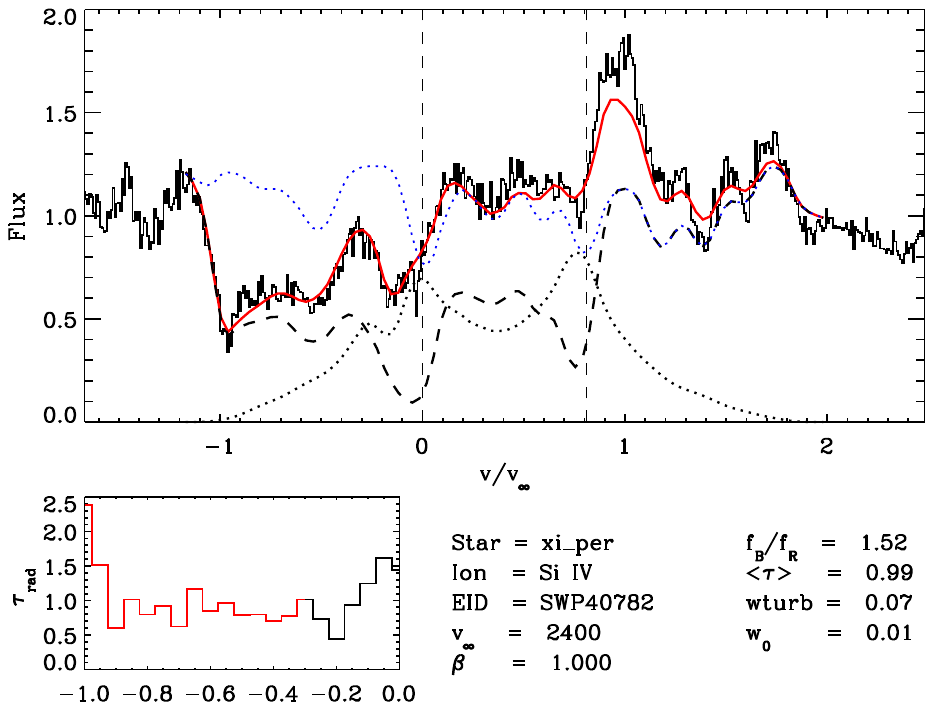} 

\vspace{-2.9in} 
\includegraphics[width=.6\linewidth]{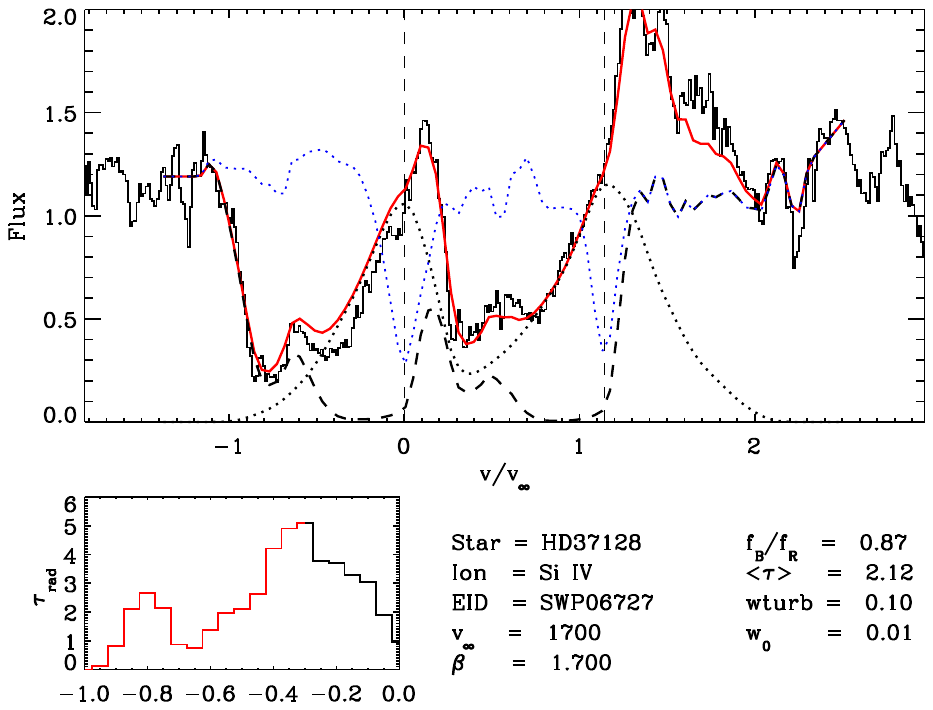}\hspace{-0.7in}\includegraphics[
width=.6\linewidth]{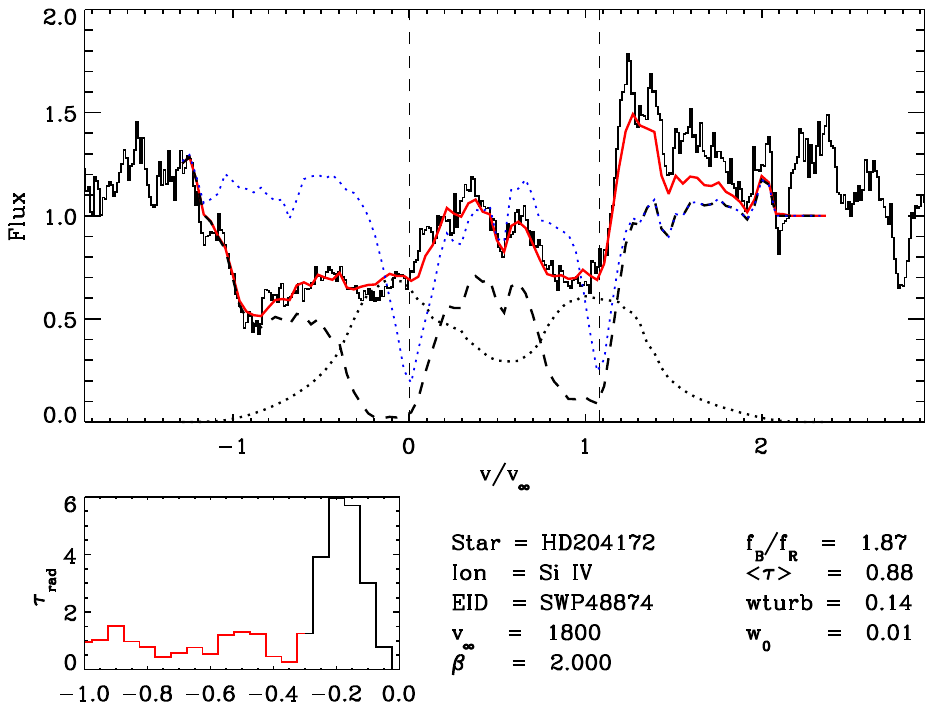} 

\vspace{-2.9in} 
\includegraphics[width=0.6\linewidth]{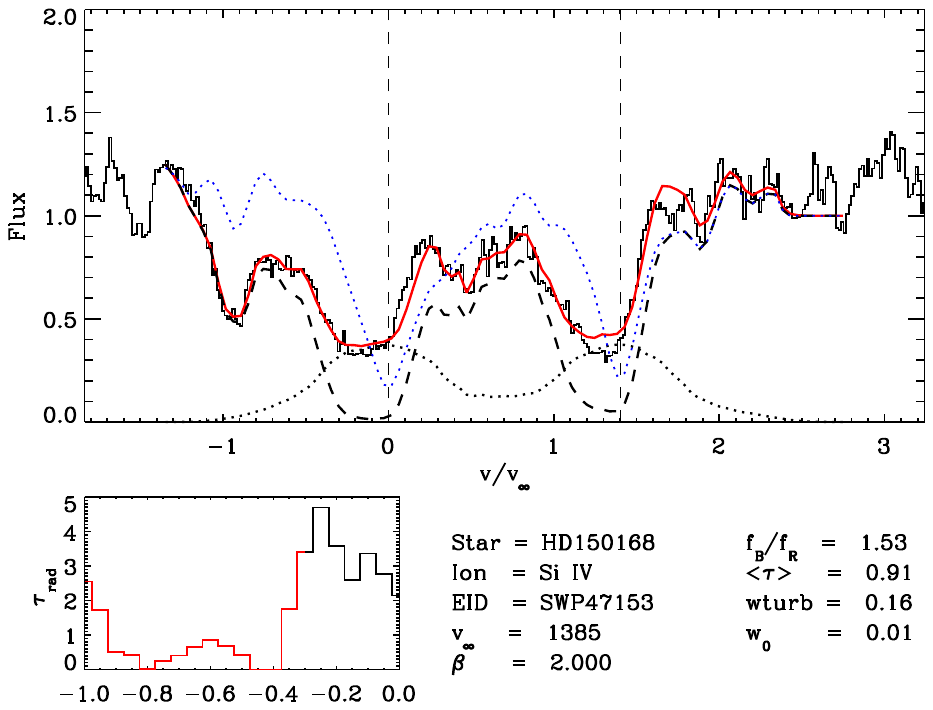}\hspace{-0.7in}\includegraphics[
width=0.6\linewidth]{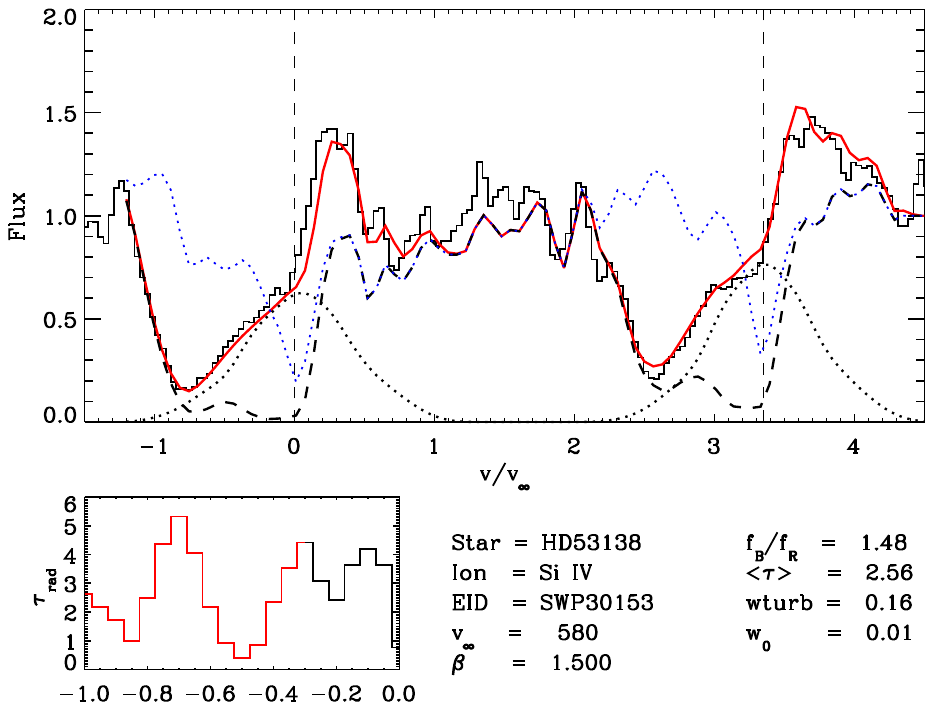} 

\vspace{-2.5in}
\caption{Examples of fits for a range of spectral types.  Each panel lists 
the name of the star, the \iue\ spectrum fit and the fit parameters (where 
wturb $\equiv w_D$).  The figures show the observed spectrum (solid black), 
the photospheric spectrum used (bue dotted), the fit (solid red), the 
emission contribution (black dotted), the absorption component (dashed 
black) and the rest wavelengths of the doublet (vertical dashed).  The 
insert is \taurad\ versus $v/v_\infty$, where the red portion is that used to obtain 
the $\tau_i^{(n)}$\ measurements.
}
\label{fig:fits}
\end{figure*}

Repeating this process, we fit 1713 spectra of the 25 stars listed in 
Table~\ref{tab:stars}.  Each fit results in a \taurad, $f_B/f_R$\ and 
$w_D$\ for that spectrum.  Fourteen of the fits ($ < 1$\%) were very poor 
for various reasons and rejected.  This left 1699 fits for analysis.  
Figure~\ref{fig:fits} shows examples of the fits for a range of 
temperatures.  When examining the fits, recall that the emission component 
is given a low weight since it originates throughout the wind.   

\section{Analysis}
In this section we first examine the time dependence of some properties of 
the fits for stars which have time series.  Next, we examine the parameters 
that characterize the mean results of the fits for all of the stars, in 
order to search for any relationships that may emerge when the entire 
sample is considered.  But first, we must define some notation.

\subsection{Notation}\label{sec:not}

We label a parameter, say $x$, derived from the $i$-th observation of the 
$n$-th star in sample as $x^{(n)}_i$.  Each fit produces 22 parameters: 20 
\taurad\ bins, $f_B/f_R$, and $w_D$.  With so many spectra to analyze, we 
distilled the \taurad\ data into a single  quantity.  For each spectrum, 
we find the mean of $\tau_{rad}(w)$\ over the normalized velocity range 
$w_1 \leq w \leq w_2$.  We use $w_1 = 0.3$ and $w_2 = 1.0$.  We adopt 0.3 
for $w_1$, because the optical depth at low $w$\ can be strongly influenced 
by the exact nature of the photospheric profile when the wind absorption is 
not very strong.  Specifically, the averaging for the $i$-th spectrum of 
the $n$-th star is defined as
\begin{equation}
\tau^{(n)}_i = \langle \tau_{rad}(w_1 < w < w_2)^{(n)}_i \rangle
\end{equation} 
To simplify the notation, we define \rni\ as the ratio of the blue and 
red $f$-values of the doublet determined by the 
fit, i.e., \begin{equation}
r_i^{(n)} \equiv f_{Bi}^{(n)}/f_{Ri}^{(n)} 
\end{equation}
Thus, each of the 1699 spectra are characterized by the following three 
parameters: \tauni, \rni, and \wdni.

In addition, we summarize all of the parameters for the $n$-th star in 
terms of their mean and standard deviations with the following notation: 
$\tau^{(n)}$, $\sigma(\tau^{(n)})$, $r^{(n)}, \; \sigma(r^{(n)})$, 
and $w_D^{(n)}, \; \sigma(w_D^{(n)})$.  We also examined the cross 
correlation coefficients of all of the variables, but only the one 
relating \tauni\ and \rni\ proved to be of interest.  It is denoted as 
$\rho(\tau^{(n)}_i, r^{(n)}_i)^{(n)}$.  Finally, we note that we often 
omit the superscript $(n)$\ when it is obvious that we are referring 
to a specific star.  

\subsection{Errors in the model parameters}\label{sec:errors}
The errors affecting the model parameters were estimated using a Monte 
Carlo procedure that is described in the Appendix.  The results are listed 
in Table~\ref{tab:errors}, where we see that the relative error ranges 
are: $0.02 \leq \sigma(r)/r \leq 0.08$, $0.02 \leq \sigma(w_D)/w_D 
\leq 0.14$, and $0.04 \leq \sigma(\tau)/\tau \leq 0.10$, with an RMS of 
0.059.  These results show that the estimates of our model parameters are 
quite robust. 

\begin{figure*}
\begin{center}
\includegraphics[width=0.5\linewidth]{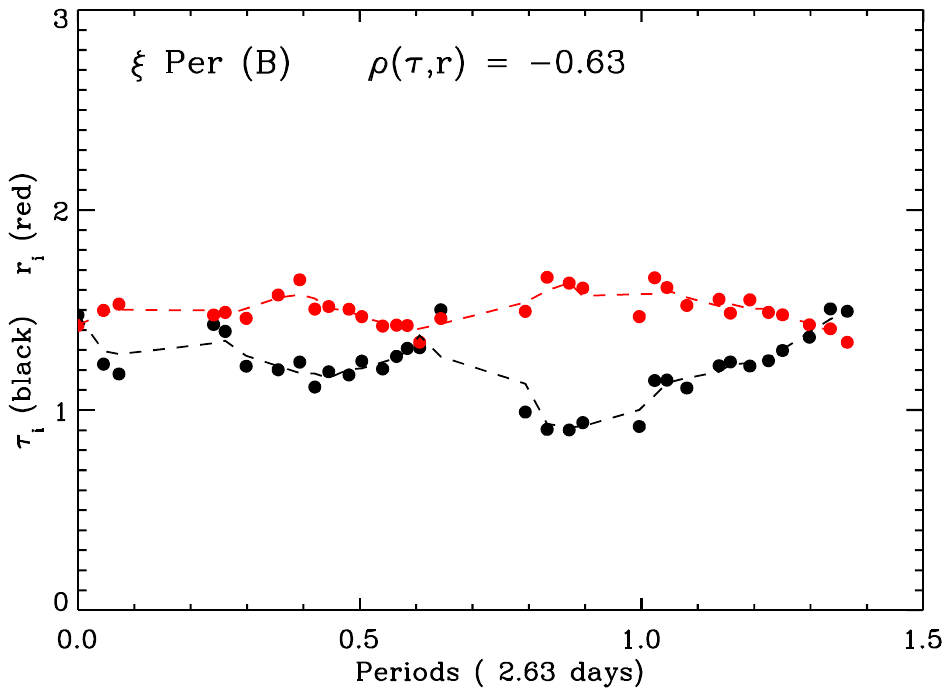}\hspace{-.5in}
\includegraphics[width=0.5\linewidth]{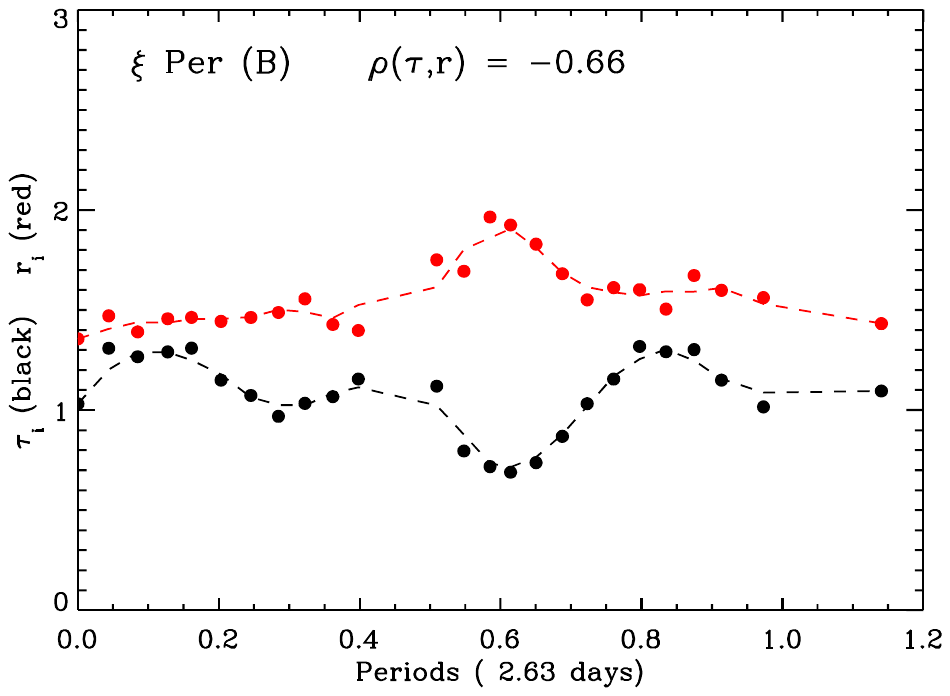} 

\vspace{-2.5in} 
\includegraphics[width=0.5\linewidth]{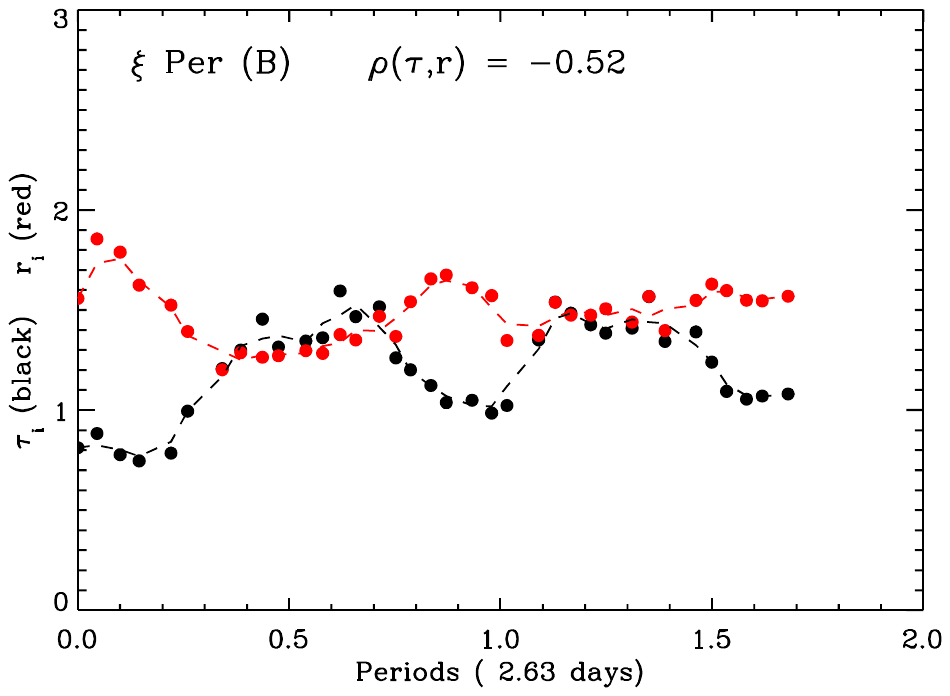}\hspace{-.5in}
\includegraphics[width=0.5\linewidth]{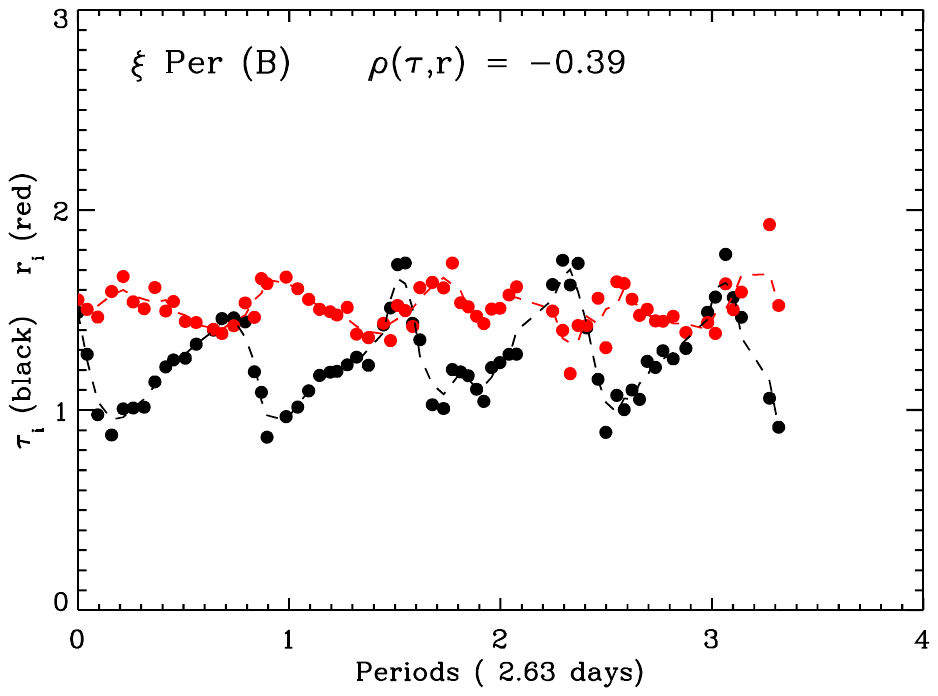} 

\vspace{-2.5in} 
\includegraphics[width=0.5\linewidth]{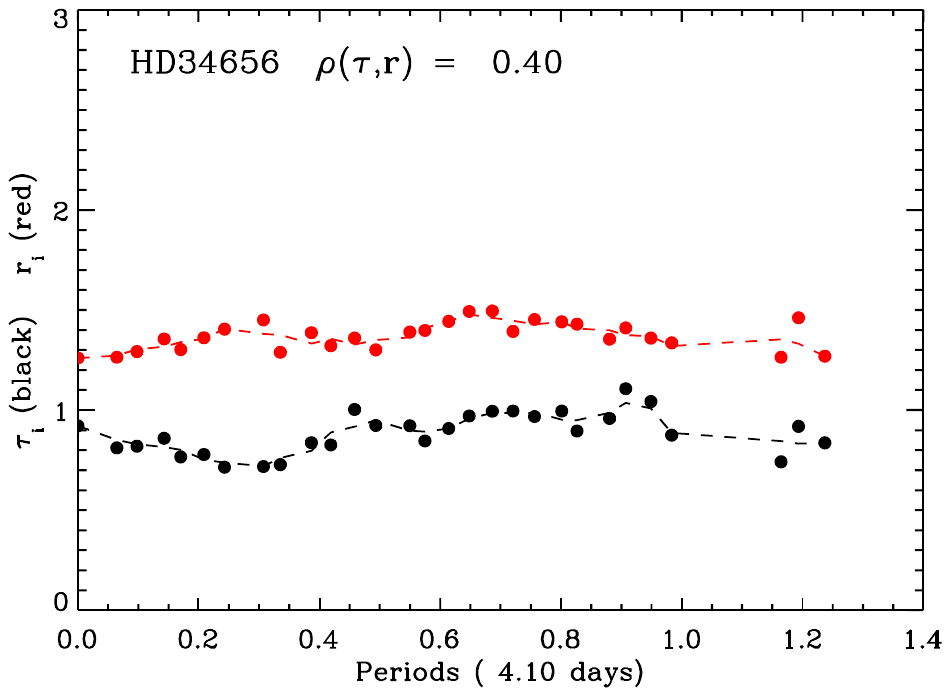}\hspace{-.5in}
\includegraphics[width=0.5\linewidth]{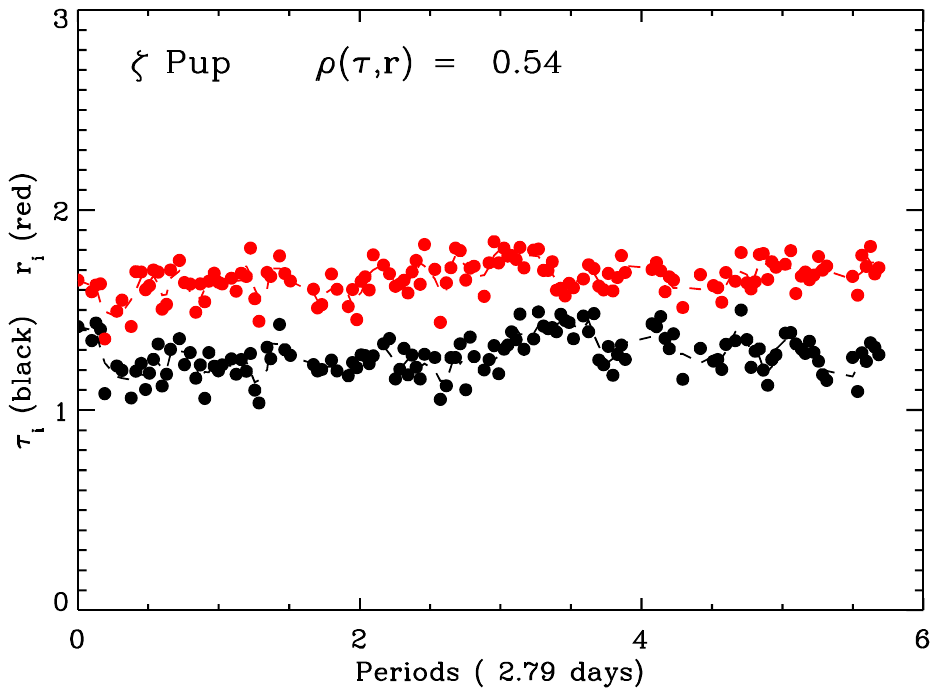} 

\end{center}
\vspace{-2.2in}
\caption{Plots of $\tau_i$\ and $r_i$\ against time, in units of the 
maximum rotation period for stars with time series, as defined in the 
text.  The dashed lines are three point smoothed versions of the points.
The name of the star and correlation coefficient between $\tau$\ and $r$\ 
are also shown in each panel.  For stars with multiple series, the plots 
are ordered chronologically left-to-right, top-to-bottom.  Stars with a 
(B) after their name have bowed structures in their dynamic spectra.}
\label{fig:frat_time1}
\end{figure*}

\begin{figure*}
\begin{center}
\includegraphics[width=0.5\linewidth]{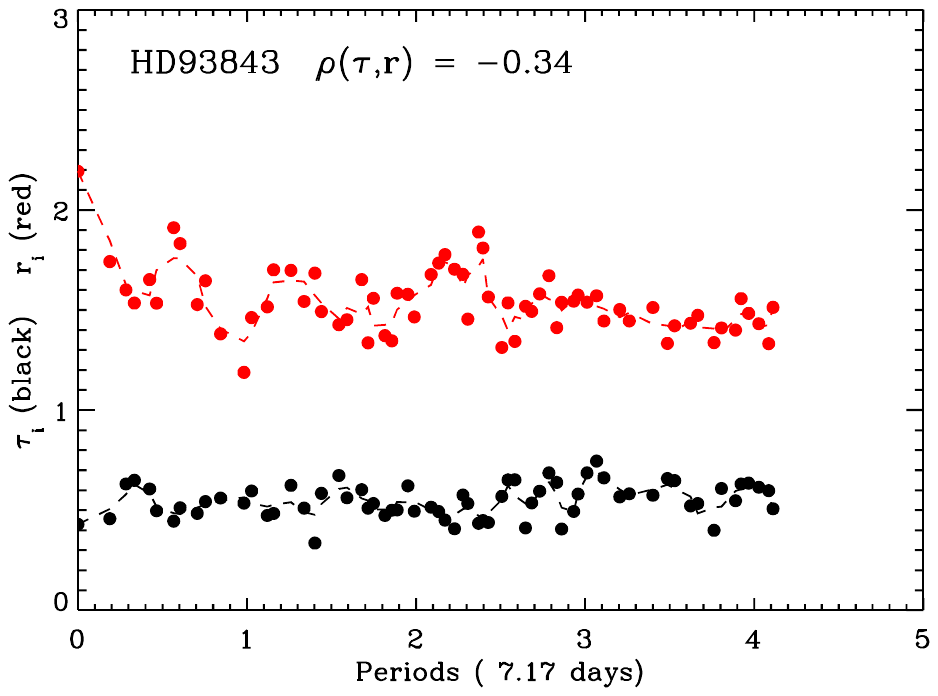}\hspace{-.5in}
\includegraphics[width=0.5\linewidth]{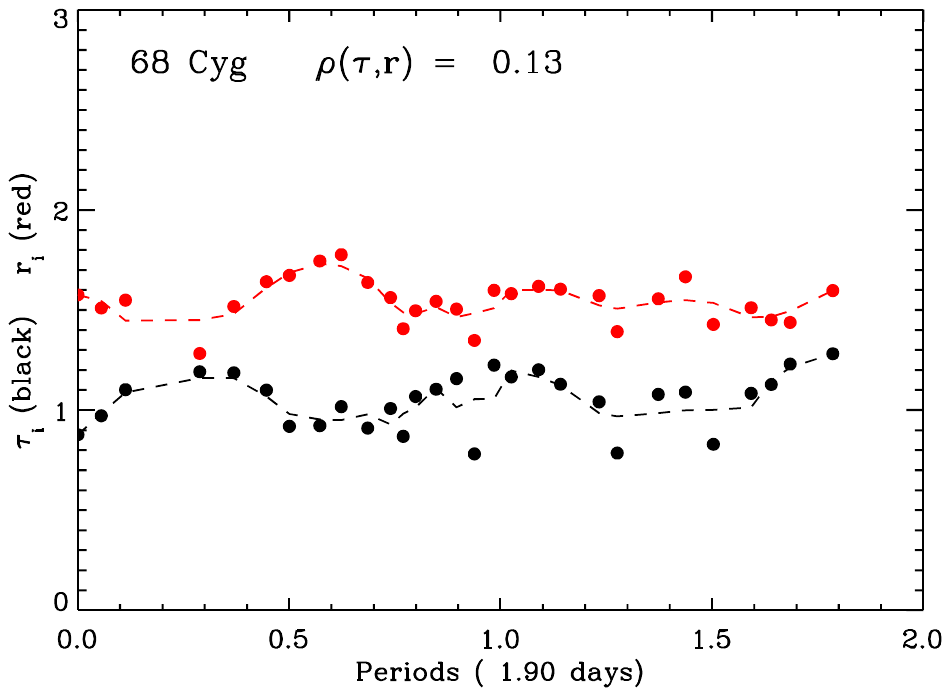} 

\vspace{-2.5in} 
\includegraphics[width=0.5\linewidth]{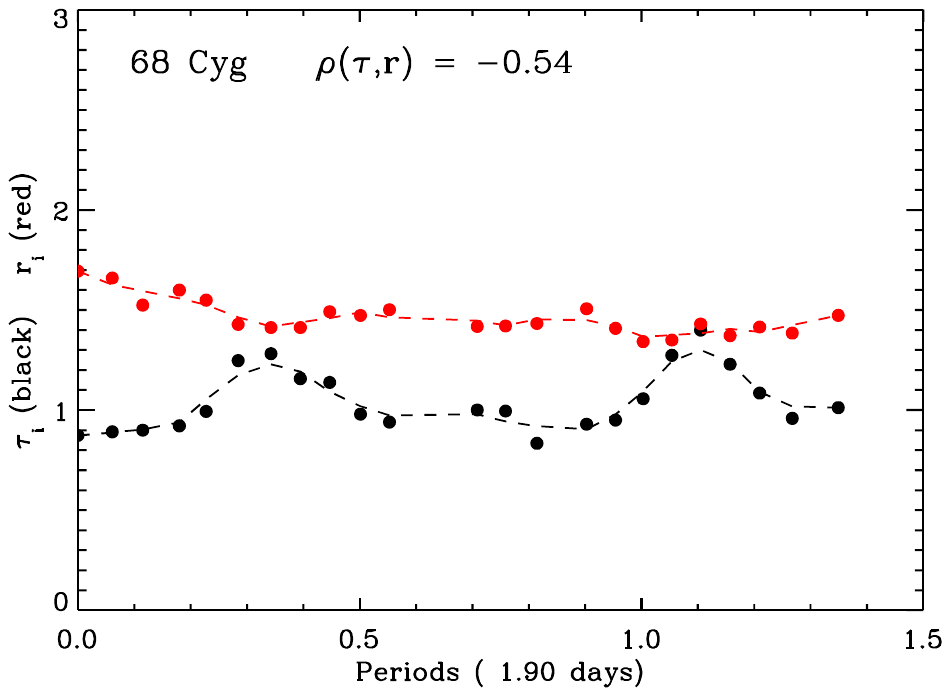}\hspace{-.5in}
\includegraphics[width=0.5\linewidth]{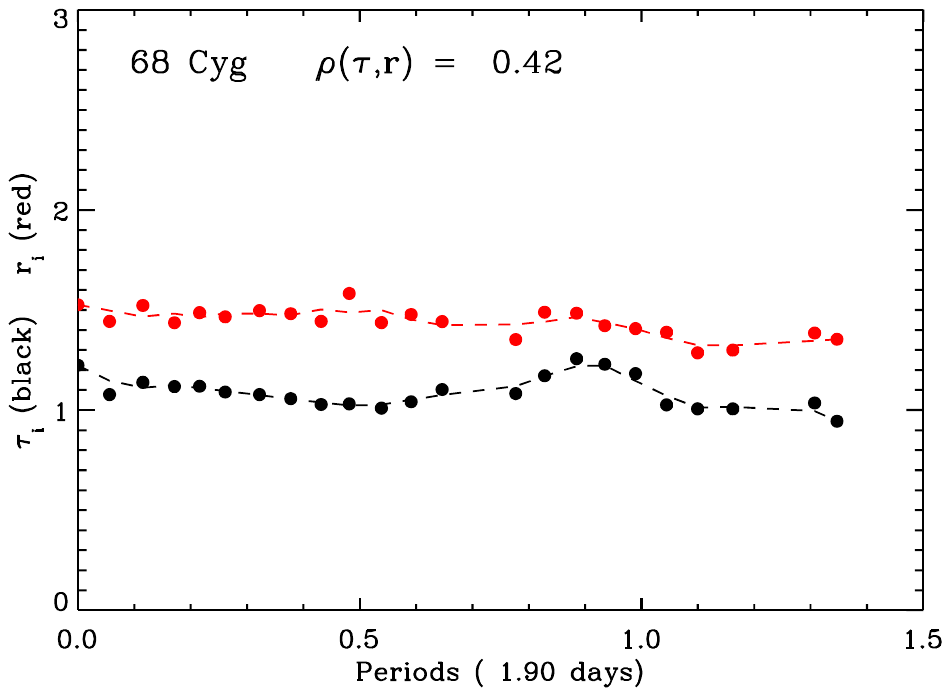} 

\vspace{-2.5in} 
\includegraphics[width=0.5\linewidth]{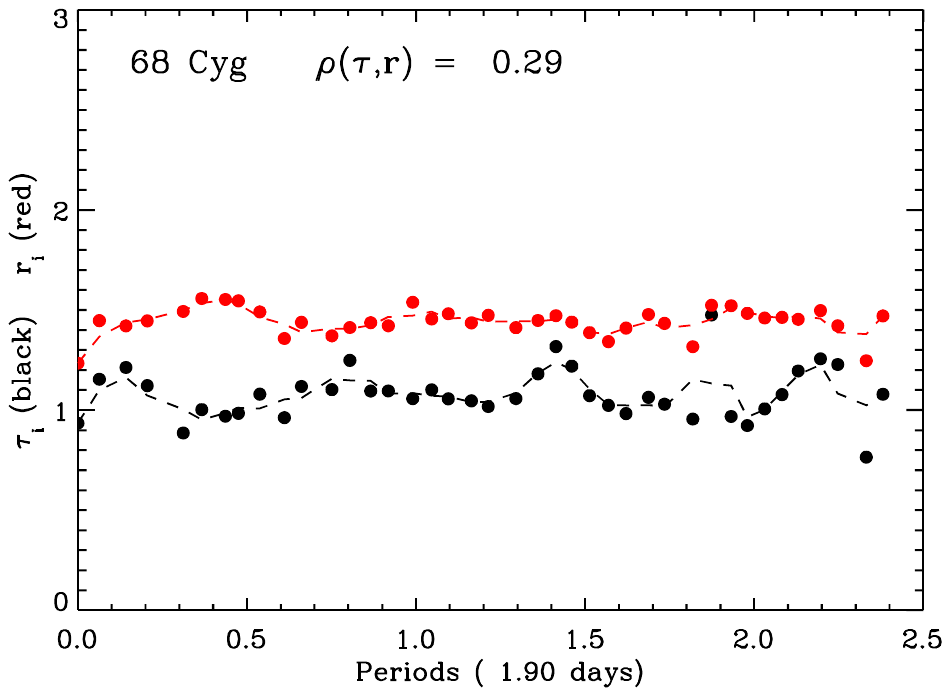}\hspace{-.5in}
\includegraphics[width=0.5\linewidth]{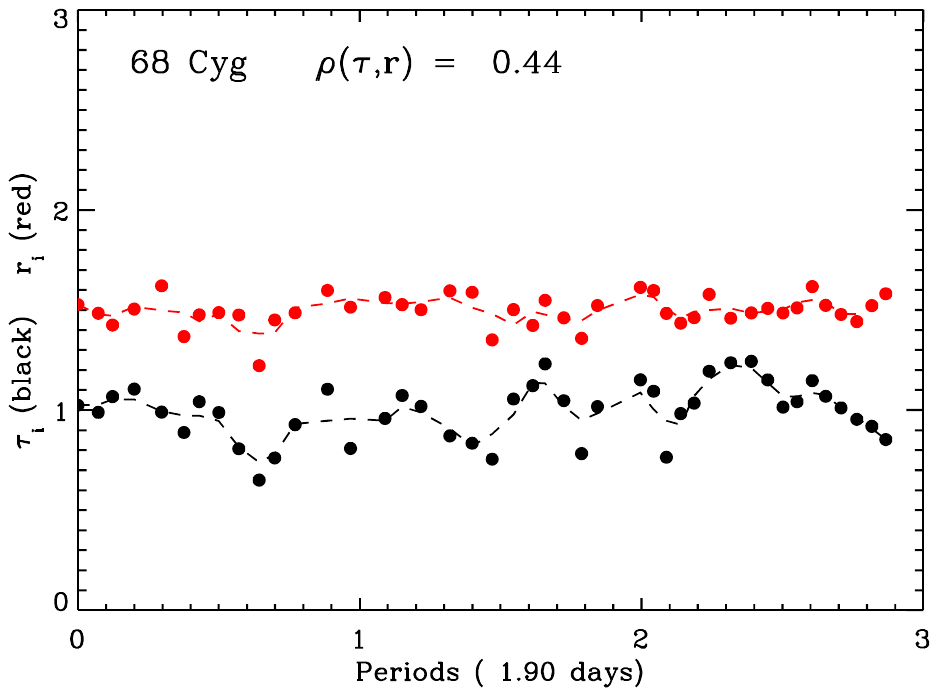} 

\end{center}
\vspace{-2.2in}
\caption{Same as Figure~\ref{fig:frat_time1}.}
\label{fig:frat_time2}
\end{figure*}

\begin{figure*}
\begin{center}
\includegraphics[width=0.5\linewidth]{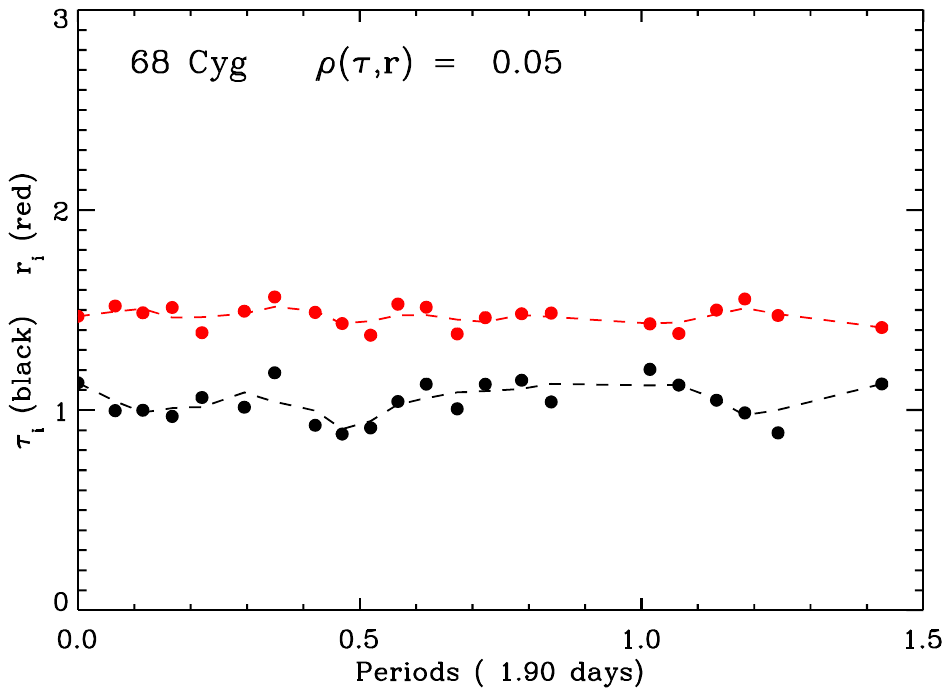}\hspace{-.5in}
\includegraphics[width=0.5\linewidth]{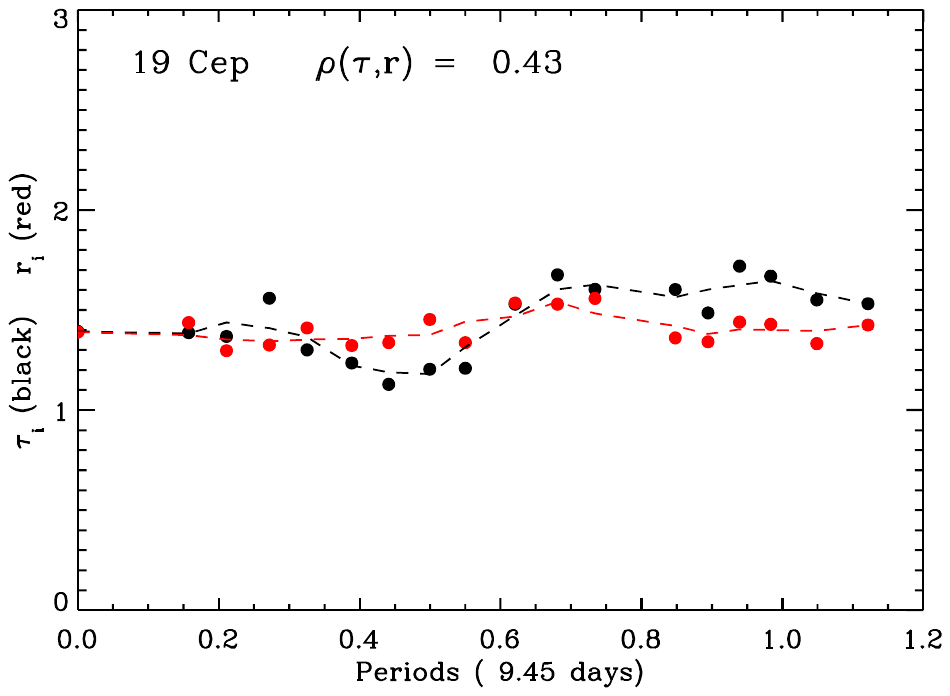} 

\vspace{-2.5in} 
\includegraphics[width=0.5\linewidth]{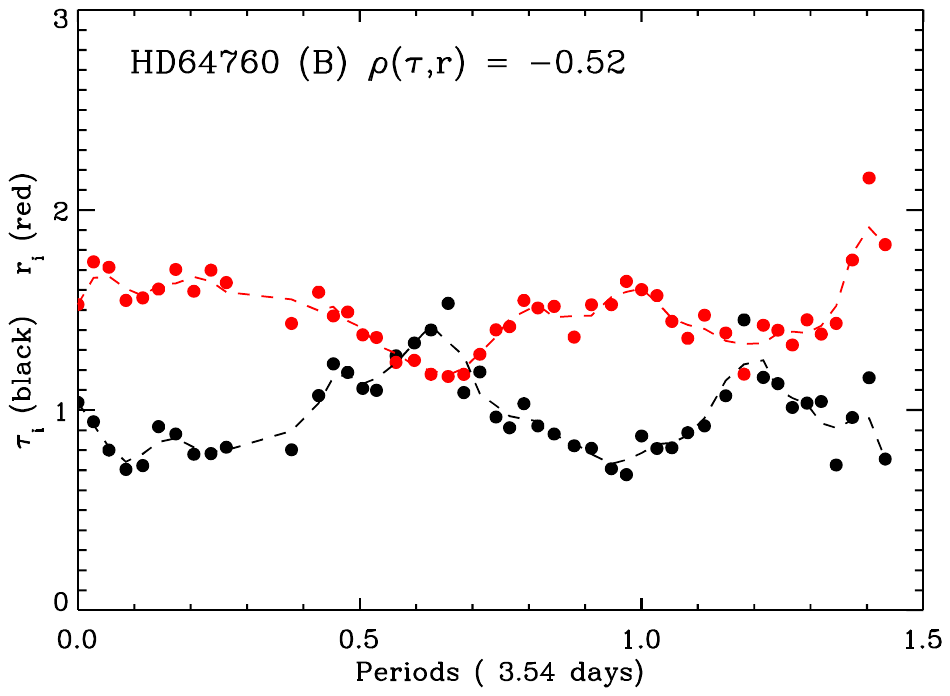}\hspace{-.5in}
\includegraphics[width=0.5\linewidth]{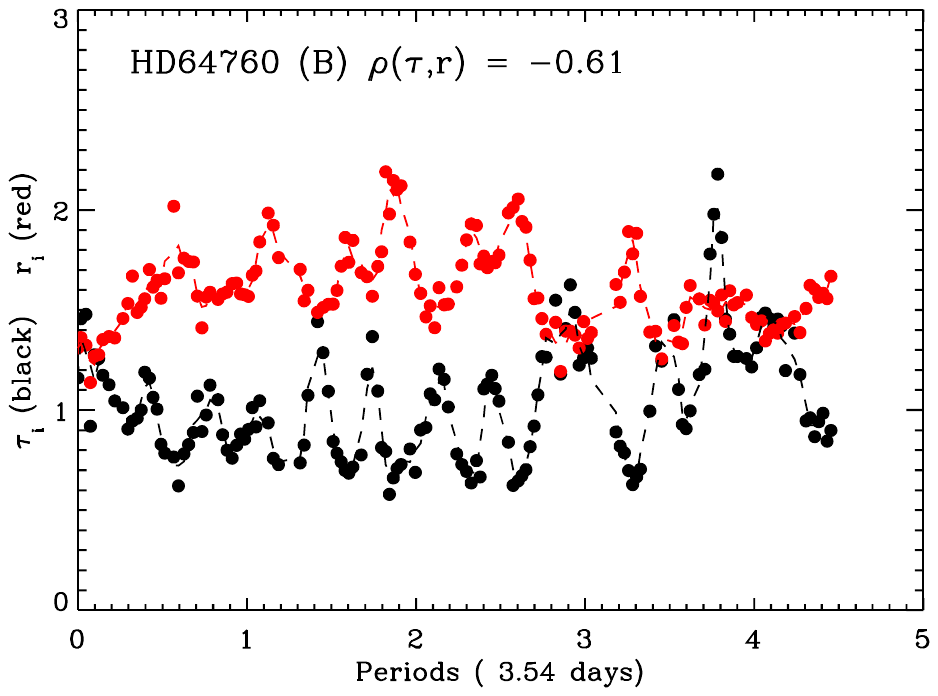} 

\vspace{-2.5in} 
\includegraphics[width=0.5\linewidth]{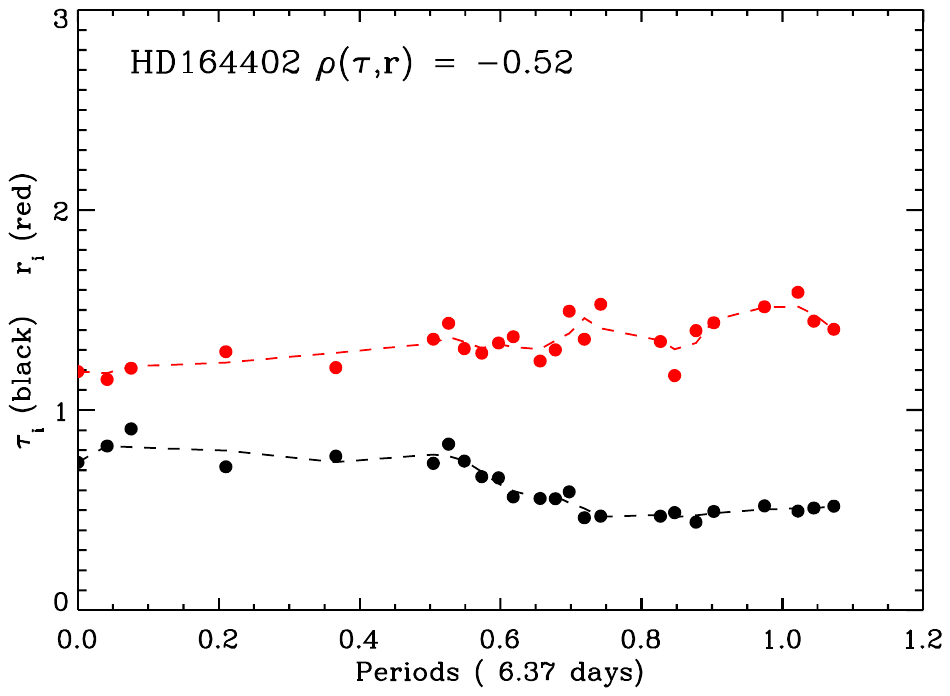}\hspace{-.5in}
\includegraphics[width=0.5\linewidth]{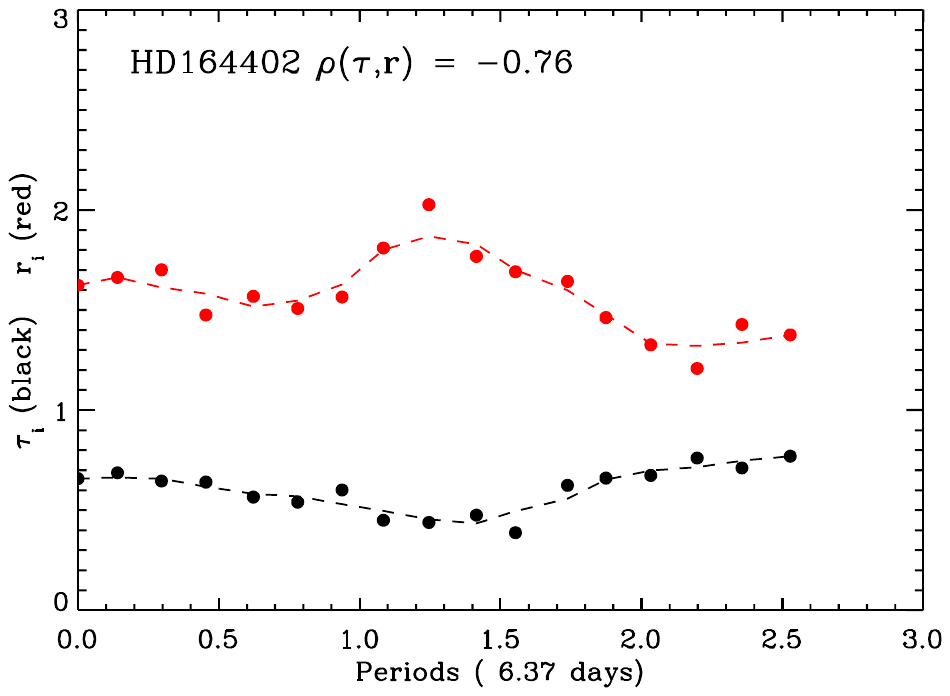} 

\end{center}
\vspace{-2.2in}
\caption{Same as Figure~\ref{fig:frat_time1}.}
\label{fig:frat_time3}
\end{figure*}

\begin{figure*}
\begin{center}
\includegraphics[width=0.5\linewidth]{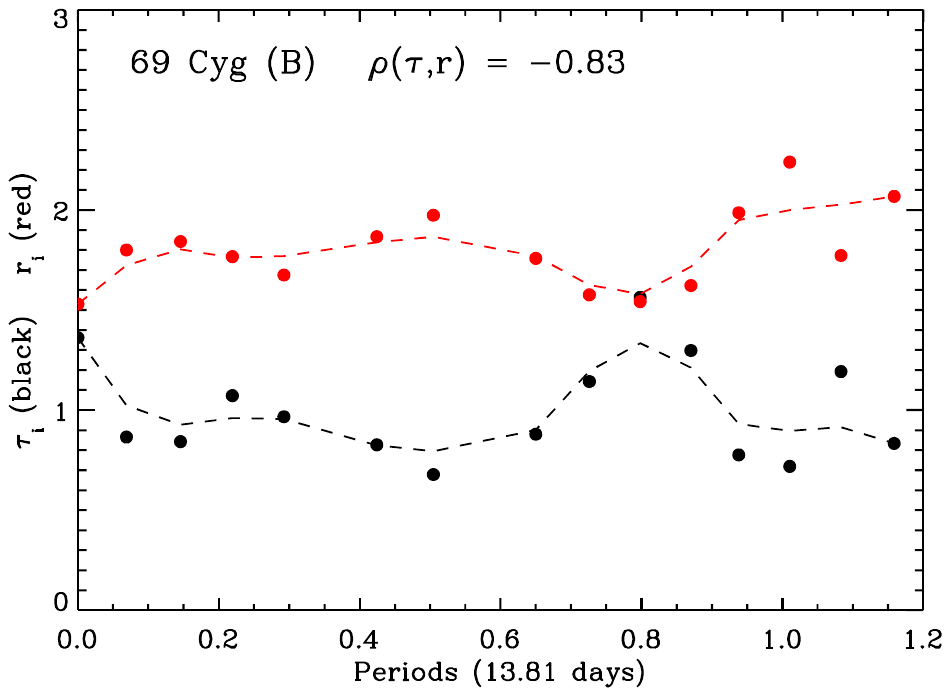}\hspace{-.5in}

\end{center}
\vspace{-2.3in}
\caption{Same as Figure~\ref{fig:frat_time1}}
\label{fig:frat_time4}
\end{figure*}

\subsection{Temporal variability of the parameters}\label{sec:temp}

Although our primary goal is to examine how wind line variability affects 
\mdot\ measurements, the large volume of data also presents an opportunity 
to examine how the different parameters might be related and how they 
behave as functions of time.  Because it is known that wind lines change 
on time scales related to the stellar rotation period \citep{prinja88}, 
we concentrate on stars with 10 or more observations spanning at least one 
rotation period.  Further, we only consider a set of observations as a 
series if it does not contain gaps larger than 20\% of the period.  Our 
search resulted in 19 series for 9 stars.  These series are summarized 
in Table~\ref{tab:series}, which lists the star names, number of spectra 
in the series, the time spanned by the series and the date of the \iue\ 
observations, and the $\tau$, $r$\ and $\rho(\tau, r)$ for each series.  

Figures~\ref{fig:frat_time1} -- \ref{fig:frat_time4} are plots of $\tau_i$\ 
and $r_i$\ versus time (in periods) for the time series.  Some stars have 
more than one series and most of these data sets are well known and have 
been described in detail before.  Multiple plots for the same star are 
presented chronologically.  Each plot gives the name of the star and 
$\rho(\tau, r)$\ for the series.  The following are descriptions of each 
series.

\medskip
\noindent {\bf \begin{boldmath}$\xi$\end{boldmath} Per:} There are four 
series that qualify for this star.  The longest one was described and 
analyzed in detail by \cite{dejong01}.  Its dynamic spectrum contains 
features termed bowed structures or bananas, which are indicative of spiral 
structures in the wind, viewed near $\sin i = 1$\ \citep{cranmer96}.   For 
this star, the variations in both $\tau_i$\ and $r_i$\ are quite strong and 
anti-correlated.  All of the series appear to contain at least one strong 
maximum per revolution period.  

\medskip
\noindent {\bf HD 34656:} There is one series for this star, and it was 
examined previously by  \cite{kaper96}.  For this star, the variations 
measured by $\tau_i$\ and $r_i$\ are not very strong and are weakly, 
positively correlated with one another.  

\medskip
\noindent {\bf \begin{boldmath}$\zeta$\end{boldmath}\ Pup:}  This star 
was part of the \iue\ MEGA campaign \citep{massa95}, and analyzed in 
detail by \cite{howarth95}.  In spite of its dynamic spectrum showing 
significant variations, the variability measured by $\tau_i$\ is quite 
weak and poorly correlated with $r_i$.  This is because the variations 
seen in dynamic spectra of $\zeta$~Pup are not bowed.  As a result, at 
any given time, there are positive and negative variations present in 
the profiles, and our broad band measure of optical depth can suppress 
such variations.  

\medskip
\noindent {\bf HD 93843:} \cite{prinja98} presented a dynamic spectrum 
of the one series for this star and analyzed it in detail. Its dynamic 
spectrum shows no sign of bowed structures.  The variations in $\tau_i$\ 
are not very strong and are weakly anti-correlated with $r_i$.  

\medskip
\noindent {\bf 68 Cyg:} There are six series for this star, and the 
dynamic spectra have been described by \cite{kaper96} and \cite{massa15}.  
The last two series are very close in time and are described as a single 
series by \cite{massa15}.  A particularly interesting aspect of these 
series is that $\tau$\ and $r$\ are strongly anti-correlated at some 
epochs, and correlated at others.  This change of behavior might be related 
to the fact that this star has the largest \vsini\ of all the stars listed 
in Table~\ref{tab:stars}.  In fact, recently \cite{brit23} measured a 
\vsini\ of 323 \kms\ for this star.  It is possible that the spiral 
structures have difficulty becoming organized or are easily disrupted by 
the rapid rotation.

\medskip
\noindent {\bf 19 Cep:} \cite{kaper96} discussed the available data for 
this star.  They presented five series, but only one has a time span 
covering more than an entire rotation period.  None of the dynamic spectra 
indicate the presence of bowed structures.  The $\tau_i$\ variations are 
not very strong and are positively correlated with those in $r_i$.


\medskip
\noindent {\bf HD 64760:} There are two series for this star and 
\cite{fullerton97} analyzed them in detail.  The dynamic spectra show 
the signatures of bananas, and were analyzed in terms of the co-rotating 
interaction region (CIR) model, but any process that creates a spiral 
pattern in the wind will do.  As pointed out by \cite{fullerton97}, 
features in the shorter, 1993 series, seem to repeat once per rotation 
while those in the 1995 series repeat twice per revolution.  In both 
series, variations in $\tau_i$\ and $r_i$\ are strong and highly 
anti-correlated.  

\medskip
\noindent {\bf HD 164402:} This star has two series, previously described 
by \cite{prinja02}.  Both plots seem to contain a broad feature which 
spans more than a rotation period.  However, one must remember that the 
maximum period uses the adopted value $v\sin i = 75$\ \kms.  If this is 
too large by only about 20 \kms, then the variation would fit into a 
single period.  In any case, variations in $\tau_i$\ and $r_i$\ are 
strongly anti-correlated, even though there is no strong signature of 
bowed structure in the available dynamic spectra.  

\medskip
\noindent {\bf 69 Cyg:} \cite{prinja02} presented the dynamic spectrum for 
this star.  Although not flagged as having a bowed structure, the dynamic 
spectra do show the broad band variations that are typical of such 
features.  Figure~\ref{fig:frat_time4}, shows a strong anti-correlation 
between $\tau_i$\ and $r_i$.    

\medskip
We note that stars whose dynamic spectra show bow shaped structures, or 
"bananas" tend to have two distinctive features in the plots shown in 
Figs.~\ref{fig:frat_time1} -- \ref{fig:frat_time4}.  The first is that 
the variability is considerably larger than those without bananas.  This 
is at least partly a natural consequence of the broad band definition of 
the $\tau_i^{(n)}$, since the bananas are also broad band features.  The 
second is that $r_i$\ and $\tau_i$\ are generally strongly anti-correlated 
for stars with bowed structure in their dynamic spectra.  Even HD~150168, 
which was identified by \cite{prinja02} as having such structure, but is 
not included because its series only spans 0.85 periods, also has a 
large variance in $\tau_i$\ which is anti-correlated with $r_i$.  

We also examined the time dependence of the parameter $w_{Di}$, which also 
varies smoothly as a function of time.  Usually it is not obviously 
related to either $\tau_i$\ or $r_i$.  The two exceptions are stars with 
strong bananas: $\xi$~Per and HD~64760.  In these cases, $w_{Di}$\ varies 
in phase with $\tau_i$. 

\subsection{Global properties of the variability}

In dealing with the entire sample of stars listed in Table~\ref{tab:stars}, 
we examine the parameters, $\tau\ {\rm and}\ \sigma(\tau)$, 
$r\ {\rm and}\ \sigma(r)$, and $w_D\ {\rm and}\ \sigma(w_D)$\ defined in 
section~\ref{sec:not}. These are just the means and standard deviations 
of the $\tau_i$, $r_i$\ and $w_{Di}$\ determined from the fits to each 
spectrum for a particular star.  We also examined the cross correlation 
coefficients between the different variables, and only $\rho(\tau, r)$ 
showed any systematic behavior.  The results are listed in 
Table~\ref{tab:taus} which gives the star name, $r$, $\sigma(r)$, $w_D$, 
$\sigma(w_D)$, $\tau$, $\sigma(\tau)$, $\sigma(\tau)/\tau$\ and 
$\rho(\tau,r)$.

\begin{figure*}
\includegraphics[width=0.55\linewidth]{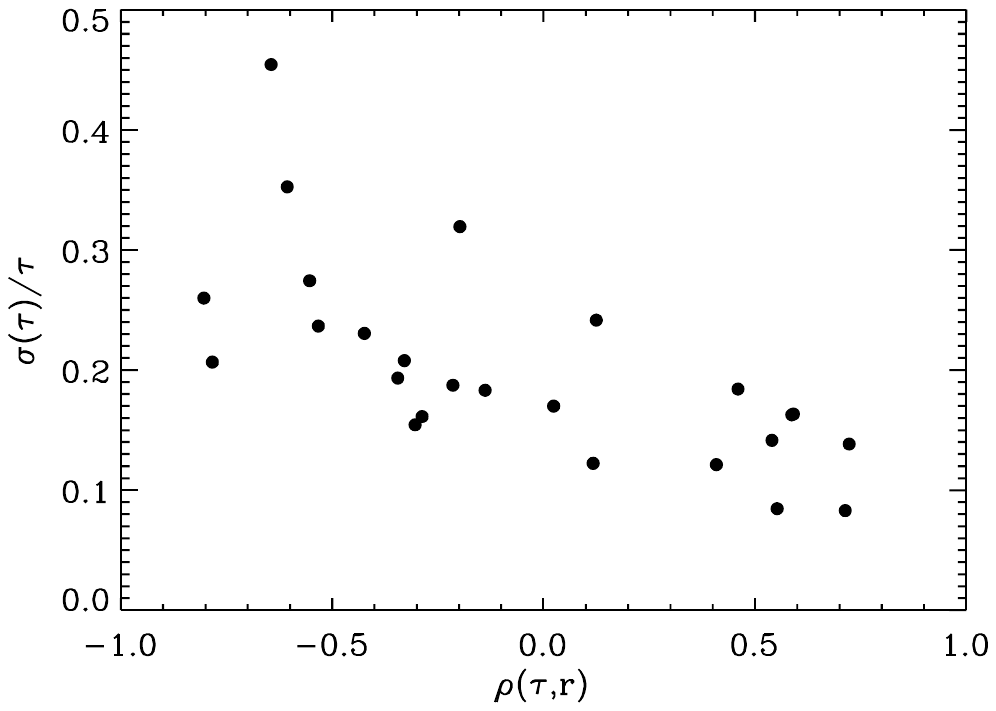}\hspace{-0.4in}
\includegraphics[width=0.55\linewidth]{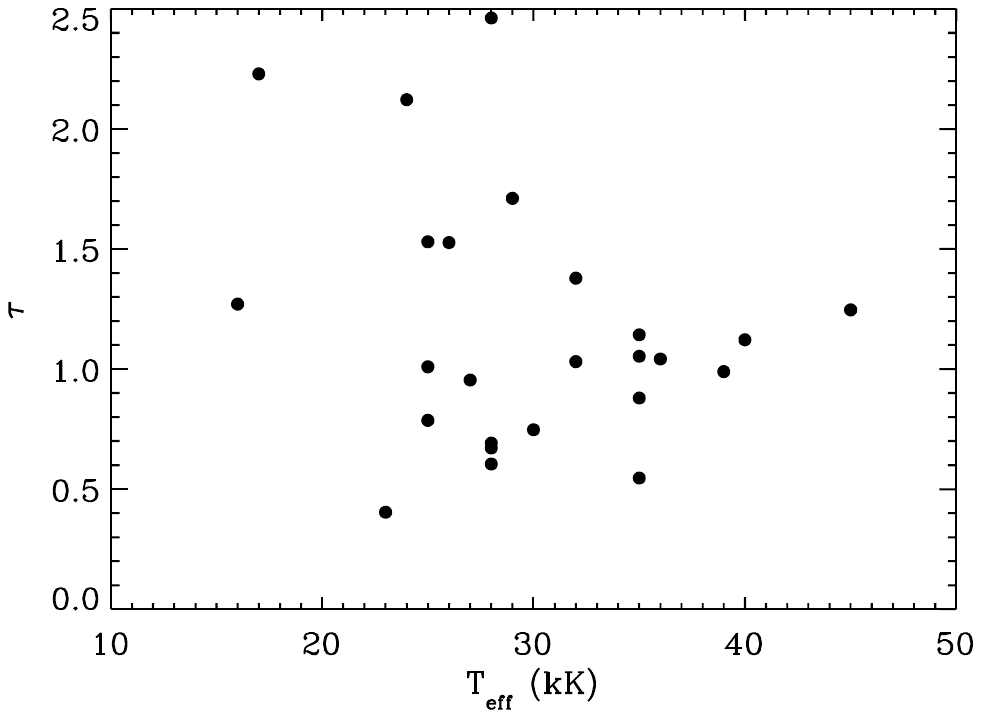}\vfill \vspace{-2.7in}
\includegraphics[width=0.55\linewidth]{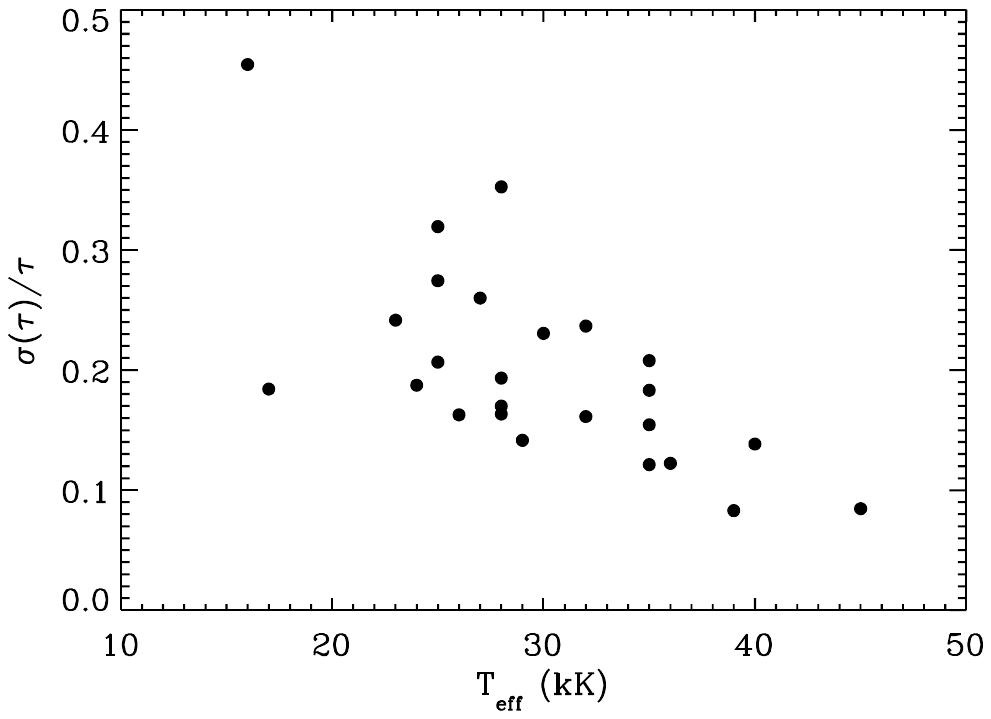}\hspace{-0.4in}
\includegraphics[width=0.55\linewidth]{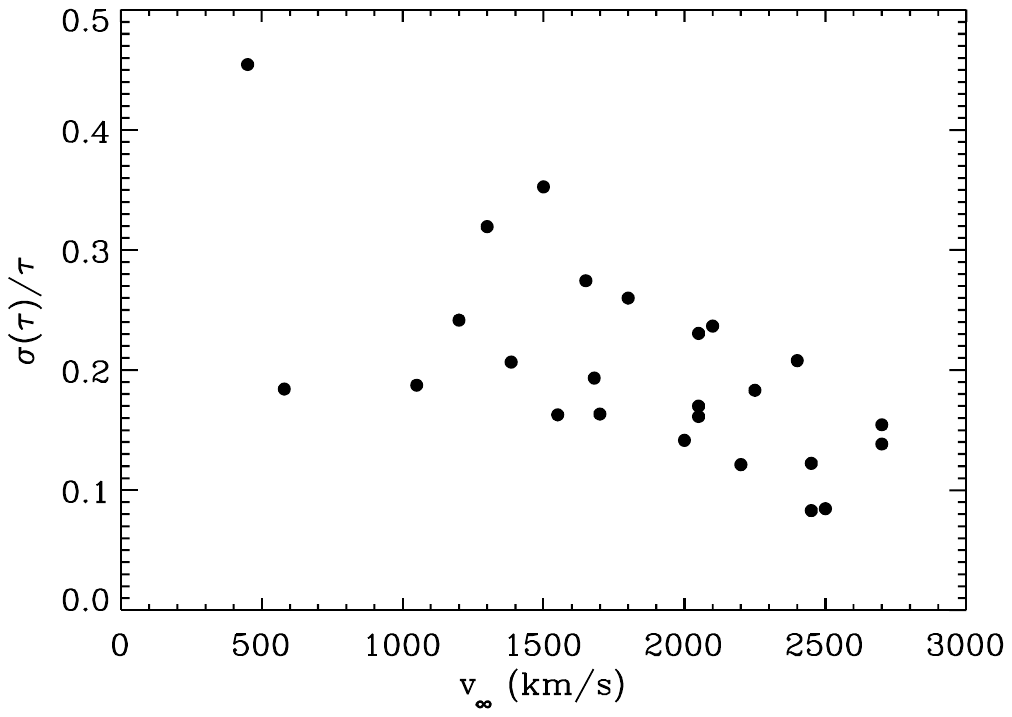}\vfill \vspace{-2.7in}
\includegraphics[width=0.55\linewidth]{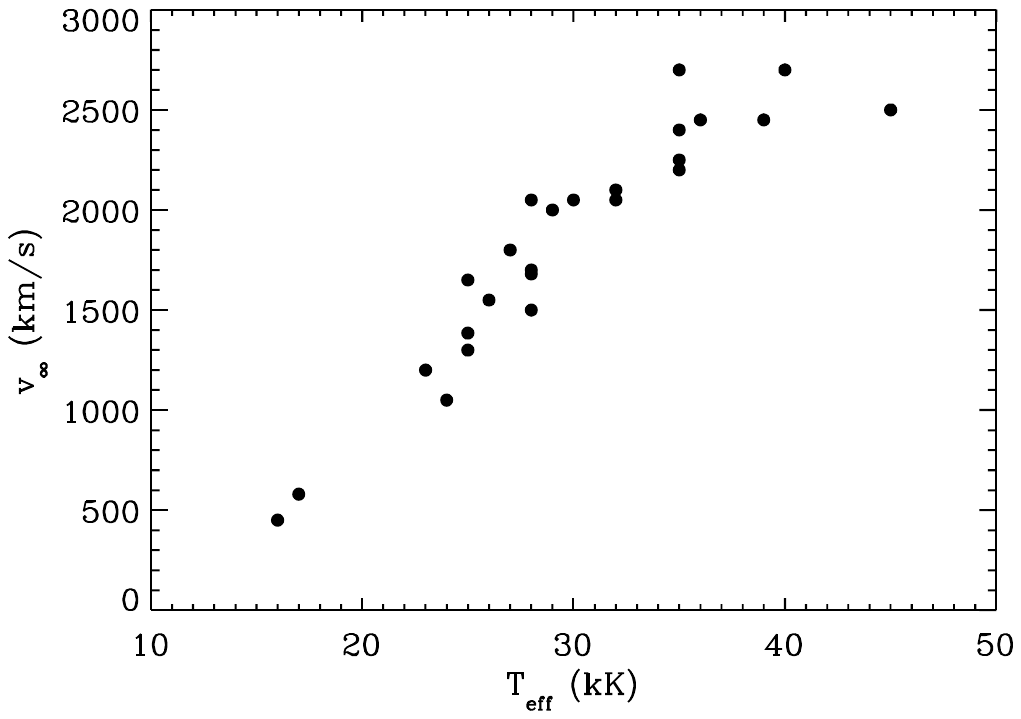}\hspace{-0.4in}
\includegraphics[width=0.55\linewidth]{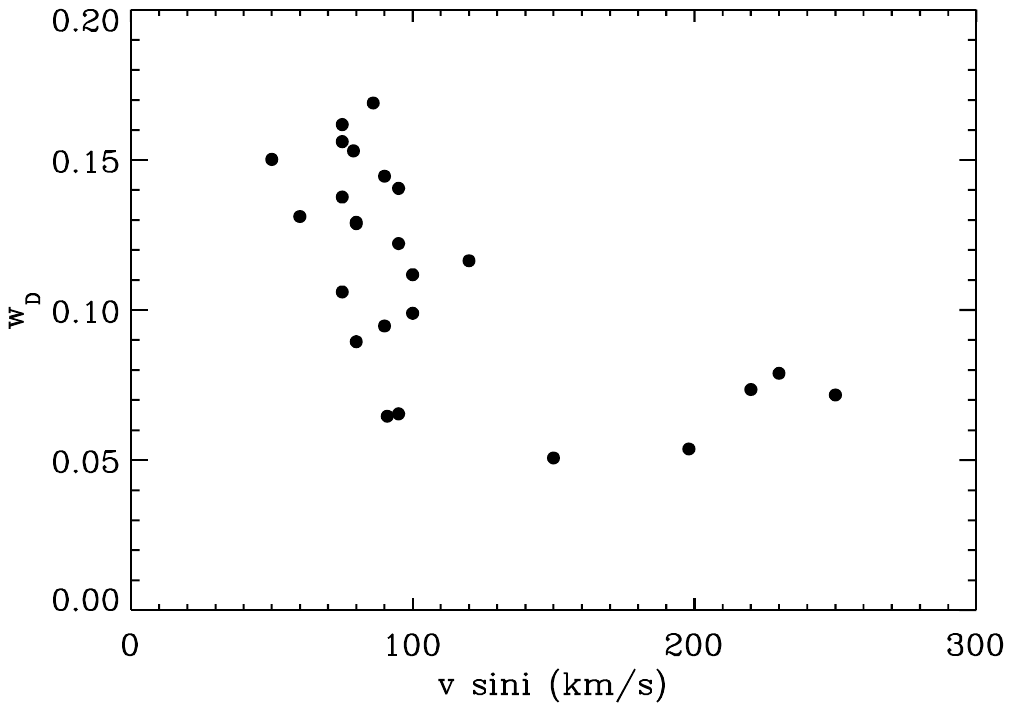}

\vspace{-2.2in}
\caption{Plots of various parameters against each other.  Top left, 
$\sigma(\tau)/\tau$\ versus the correlation coefficient between $\tau$\ and 
$r$.  Top right, $\tau$\ versus \teff.  Middle left, $\sigma(\tau)/\tau$\ 
versus \teff.  Middle right, $\sigma(\tau)/\tau$\ versus \vinf.  Lower left, 
\vinf\ versus \teff. Lower right, $w_D$\ versus \vsini.  
 }
\label{fig:some_plots}
\end{figure*}

Because one of our main goals is to quantify relative errors in $\dot{M}$, 
the important parameter to consider is $\sigma(\tau)/\tau$, since it is 
proportional to $\sigma(\dot{M})/\dot{M}$.  We examined all permutations 
of the variables described above, and selected results are shown in 
Figure~\ref{fig:some_plots}.  The only fit parameter that shows a 
significant correlation with $\sigma(\tau)/\tau$\ is $\rho(\tau, r)$, and 
this is shown in the upper left of fig.~\ref{fig:some_plots}.  This 
reflects what was seen in the temporal plots. There, we noted that the 
stars whose dynamic spectra showed bowed structure also had large 
variations in $\tau_i$, and that $r_i$\ and $\tau_i$\ tended to be 
anti-correlated, $\rho(\tau, r) < 0$.  

In addition to the wind line measurements, we also have the following 
ancillary data at our disposal:  $T_{eff}$, $v \sin i$, \vinf, and 
$P_{max}$.  Some of these are also shown in Figure~\ref{fig:some_plots}.  
The upper right plot shows that $\tau$\ is independent of \teff.  This 
is expected since our sample was selected to have well developed but 
unsaturated \siiv\ wind lines.  In contrast, the middle left plot shows 
that the fractional scatter increases with decreasing \teff.  The 
middle right plot shows that $\sigma(\tau)/\tau$\ is also strongly 
correlated with \vinf.  However, this arises because \vinf\ and \teff\ 
are strongly correlated for our sample (see the lower left plot).  
This correlation may arise because our sample constrains the optical 
depth of the \siiv\ $\lambda$\ 1400 \AA\ doublet to a narrow range, 
effectively confining the sample to a narrow slice across the HRD, with 
the cooler stars being more luminous.

Examination of the plots between the ancillary data and our derived 
parameters shows that there is no relation between $r$\ or $\sigma(r)$ 
and any of the ancillary data.  The same is true for $w_D$, with the 
exception of $w_D$\ and \vsini, shown in lower right plot of 
Figure~\ref{fig:some_plots}.  It suggests that $w_D$ is smaller in stars 
with larger \vsini\ values.  

\section{Discussion} \label{sec:discussion}
The discussion addresses two separate issues.  The first involves the 
implications of our results on any measurement of \mdot, and is purely 
empirical.  The second concerns how the relationships we have uncovered 
can be interpreted in physical terms.  

Throughout the discussion, we must keep in mind that our sample is biased.  
The stars were selected to have \siiv\ wind lines that are well developed, 
but unsaturated.  For the later type stars, \siiv\ may be the dominant Si 
ion in the wind.  However, for the earlier stars, it may only be a trace 
ionization stage, with most of the Si in Si~{\sc v}.  Consequently, the 
\siiv~$1400$\ doublet may be sampling very different components of the 
wind in the coolest and hottest stars in our sample.  

\subsection{Empirical implications}
\begin{figure*}
\begin{center}
\includegraphics[width=0.8\linewidth]{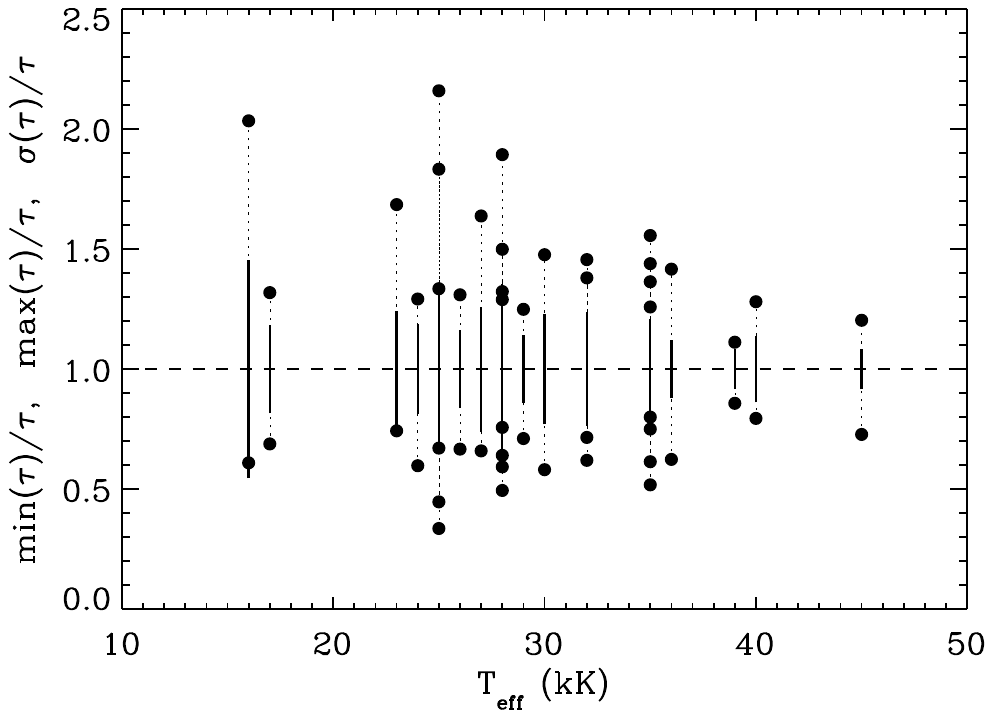}
\end{center}
\vspace{-3.5in}
\caption{The maximum and minimum values of $\tau_i$ for each star divided 
  by the mean value, $\tau$, and $\sigma(\tau)/\tau$.  There is one set of 
  points for each star.  The vertical dotted lines connect the maximum and 
  minimum values of  $\tau_i/\tau$\ and the thick bars are $\pm 1 
  \sigma(\tau)/\tau$.}
\label{fig:terrors}
\end{figure*}

In this section, we assess and quantify the intrinsic accuracy of \mdot\ 
measurements. Since \taurad\ is directly proportional to \mdot, variations 
in \taurad\ are surrogates for variations in \mdot\ measurements.  
Regardless of whether the optical depth variations are due to changes in 
the ionization fraction, $q$, or the velocity gradient, $dw/dx$, the result 
is the same: the {\em measured} \mdot\ will change proportionally.  Thus, 
for the purpose of determining the intrinsic error, it does not matter 
what causes the variations in the $\tau_i$.  

For each star, Figure~\ref{fig:terrors} shows the maximum and minimum 
measured values of $\tau_i$\ divided by the mean and $\sigma(\tau)/\tau$\ 
ploted against \teff.  These are representative of the error expected from 
a single observation.  The results imply an RMS intrinsic error between 
8\% and 45\%, with an average of 22\%. Further, the figure shows that a 
single determination can have an error as large as a factor of two or 
more.  Tere is also an indication that the errors are larger for cooler 
stars and smaller for hotter stars. 

For nearly all of the stars, the errors derived in section~\ref{sec:errors} 
are much smaller than the observed variations.  There are, however, two 
exceptions: $\zeta$~Pup and HD~190429.  As discussed in 
section~\ref{sec:temp}, our broad band measure of optical depths supressed 
the profile variations that are known to be present in $\zeta$~Pup.  
However, HD~190429 has not been closely studied for variability.  Although 
there are 16 \iue\ observations of it, these effectively sample only 4 
independent epochs.  There is one observation in 1979, thirteen in 1990 
(all obtained within 2 days), one in May of 1995 and the last in September 
of 1995.  Close examination of the 4 independent epochs shows no detectable 
variations in the \siiv\ profile.  Thus, it appears that the wind lines of 
HD~190429 do not vary at a level detectable by \iue.  That makes HD~190429 
the only star in our sample whose wind lines may not vary.  

Overall, the current discussion suggests that one should not place too much 
weight on a single observational determination of \mdot, and that two or 
more observations, well separated in time, are much more reliable.  

\subsection{Physical implications}

In this section, we examine our results for clues that may provide a deeper 
understanding of the physical processes responsible for the variability.  
It is important to understand the geometry of the structures causing the 
optical depth variations, since radiation transfer through optically thick 
structures depends on the geometry.  

We interpret our results in terms of the simple model proposed by 
\cite{cranmer96}, which predicts that spiral structures in the wind 
are responsible for the variability.  In addition, we also assume that the 
spiral structures are optically very thick, in accordance with the results 
of \cite{prinja10}.  

In this model, the bowed structures in the dynamic spectra of stars viewed 
with $\sin i \simeq 1$\ result from the way a spiral arm exits the line of 
sight to the stellar disk.  The portion that exits first absorbs at over 
a large range of intermediate velocities, so the absorption lessens there 
first.  This effect is illustrated by Figure 12 in \cite{fullerton97}.  
However, when $\sin i$ is much less than 1, the portion of the arm that 
exits the line of sight first absorbs near $v = 0$, and the effect is lost.  
Thus, we assume that although spiral structures are responsible for most, 
or possibly all, of the variability, stars whose dynamic spectra contain 
bowed structures have $\sin i \simeq 1$.  

With this model in mind, we note that Figs.~\ref{fig:frat_time1} -- 
\ref{fig:frat_time4} shows that for stars with bowed structures in their 
dynamic spectra when $\tau_i$\ increases, $r_i$\ decreases.  This 
indicates that when the overall measured opacity is larger, so is the 
covering factor, reflected in the decrease in $r_i$.  Such an effect is 
consistent with the model, where the measured absorption is strongly 
influenced by how much of the line of sight to the stellar disk is 
obscured by the optically very thick structures in the wind.  In two 
cases ($\xi$\ Per and HD 64760) when $\tau_i$\ is large, $w_{Di}$ is 
too.  This suggests that a more turbulent flow is present when the wind 
along the line of sight is composed of two very distinct flows.  

A second interesting result is that $\sigma(\tau)/\tau$\ appears to 
decrease with increasing \teff.  We cannot be certain whether this is 
simply because none of the earliest stars happen to have $\sin i \simeq 
1$\ or that the wind structure is intrinsically different.  However the 
fact that $\zeta$\ Pup has a large \vsini\ (suggesting  its $\sin i \simeq 
1$) but does not have strongly bowed structures, bolsters the notion that 
the winds of the earliest stars are, in fact, different.  
It is quite possible that the spirals that give rise to the bowed 
structures are disrupted by the more powerful winds in the earliest stars 
and fragment into a more random distribution of clumps.  

\section{Summary} \label{sec:summary}

We reviewed the evidence that wind line variability is universal and 
probably due to large, spatially coherent spiral structures.  These 
structures extend to the base of the wind and cover much of the stellar 
disk. Further, they are denser than their surroundings and, because they 
are optically thick, {\em their geometry matters}. 

We then modeled the \siiv\ wind doublet in 1699 spectra of of 25 stars.  
This allowed us to determine $\sigma(\tau)/\tau$, which is directly 
proportional to any $\sigma(\dot{M})/\dot{M}$\ estimate derived from 
fitting the \siiv\ doublet.  For the entire sample, we found that 
$\sigma(\tau)/\tau \simeq 0.22$, implying an intrinsic error of $\simeq 
22$ \% in any \mdot\ measurement.  We also showed that it is quite possible 
that a single observation can differ from the mean by as much as a factor of 
two.  In addition, there is an indication that the magnitude of the errors 
depends on the stellar temperature, being larger for the B and late O stars 
and smaller for the hottest O stars.  

In analyzing the available time series, we discovered that, for many stars, 
variations in both the optical depths and the ratios of the $f$ values are 
cyclical and $180^\circ$\ out of phase.  This is particularly strong for 
stars whose dynamic spectra contain bowed structures.  We argue that these 
stars likely have $\sin i \simeq 1$, and that the relationship is an 
indication that the increase in measured optical depth is the result of 
optically very thick structures occulting more of the line of sight to the 
stellar disk.  We also noted that, for our sample, the variations in $\tau$\ 
are smaller for the hottest stars and they are not anti-correlated with the 
ratio of $f$ values.  This raises the possibility that the spiral structures 
are torn apart in the stars with the most powerful winds.

\begin{acknowledgments}
D.M. acknowledges support from NASA ADAP grant 80NSSC22K0492 and NASA HST 
grants HST-GO-16230.001-A and HST-GO-16182.002.  L.O. acknowledges support 
from Deutsche Forschungsgemeinschaft (DFG) grant GZ: OS 292/6-1.  We also 
thank the referee for useful comments that helped clarify the presentation.  
\end{acknowledgments}

\facilities{Simbad, IUE, MAST, GAIA}



\appendix

\section{Errors in the model parameters}\label{sec:appendix}
In this appendix, we explain how we estimated the errors affecting the 
parameters derived from the non-linear least squares fits.  We began by 
randomly selecting one spectrum for each program star, $f(\lambda)_0$.  
Together with the errors for this spectrum, $\sigma(\lambda)$\ (described 
below), we generated a wavelength array of normally distributed random 
numbers with standard deviations of $\sigma(\lambda)$.  This array, 
$\epsilon(\lambda)$, was then added to the original spectrum, 
$f(\lambda)^{(n)} = f(\lambda)_0 + \epsilon(\lambda)^{(n)}$.  The process 
was then repeated 100 times to create copies of the original spectrun 
including the effects of errors.  Each of these 100 spectra were then fit 
by our model, using the same photospheric spectrum, \vinf, and $\beta$.  
Finally, the scatter in the derived model parameters were used to estimate 
their errors. 

As mentioned above, the analysis requires the expected error at each 
wavelength, $\sigma(\lambda)$.  Unfortunately, the error arrays 
accompanying the NEWSIPS spectra are not appropriate for our analysis.  
This is because they include an off-set error due to thermal changes in 
the focus of the \iue\ telescope and imprecise centering of the 
star in the \iue\ aperture.  These errors are especially large for small 
aperture spectra, since the stellar image overfilled the small aperture.  
In contrast, our data are not affected by this error because all of the 
spectra are normalized over a spectral band near \siiv.  Consequently, we 
had to determine the errors that are appropriate for our spectra.  This 
was done using the four program stars with 200 or more observations: 
$\xi$~Per, $\zeta$~Pup, 68 Cyg, and HD 64760. For these stars. we simply 
calculated the standard deviation of the normalized spectra at each 
wavelength.  We found that for wavelengths away from the intrinsically 
variable resonance lines, the $\sigma(\lambda)$\ are proportional to the 
mean flux, regardless of whether the spectra were obtained through 
the large or small aperture.  As a result, we adopted the relation, 
$\sigma(\lambda) = a f(\lambda)_0$\ to determine the errors, with 
$a = 0.06$.  The results are presented in Table~\ref{tab:errors} which 
lists: the star name, the SWP number of the spectrum used in the analysis, 
the mean and standard deviation divided by the mean of the 100 trials for 
each of the model parameters: $\tau$, $r$\ and $w_D$.


\vspace{5mm}
\facilities{HST(MAST), \iue}

\begin{table}[h]
\caption{{\bf Program Stars }}
\begin{center}
\begin{tabular}{llrrrrl}\hline
Name      &  Sp Ty            & N   & $\Delta t$ (days) & $v\sin i$\ & 
$P_{max}$ (days) \\  \hline
$\xi$\ Per      &  O7.5 III(n)((f)) & 323  &  6036 &  230 &  2.63  \\
HD 34656        &  O7 II(f)         &  30  &  3364 &   91 &  4.10  \\
HD 36486        &  O9.5 II          &  58  &  4793 &  120 &  3.58  \\
$\zeta$\ Pup    &  O4 I(n)f         & 212  &  6271 &  198 &  2.79  \\
HD 93403        &  O5 III(fc) var   &  10  &  6050 &  100 & 12.06  \\
HD 93843        &  O5 III(fc)       &  69  &  5977 &   95 &  7.17  \\
$\mu$\ Nor      &  O9.7 Iab         &  34  &  3974 &   80 & 10.93  \\
HD 162978       &  O7.5 II((f))     &  22  &  5700 &   86 &  5.89  \\
HD 190429       &  O4 If+           &  16  &  6009 &   90 &  9.86  \\
68 Cyg          &  O7.5 III:n((f))  & 227  &  6025 &  250 &  1.90  \\
HD 207198       &  O9 Ib-II         &  13  &  1827 &   80 &  8.07  \\
19 Cep          &  O9.5 Ib          & 133  &  6024 &   95 &  9.45  \\ \hline
$\zeta$\ Per    &  B1 Ib            &  14  &   370 &   75 & 13.07  \\
$\epsilon$\ Ori &  B0 Ia            &  53  &  3080 &   75 & 22.08  \\
$\kappa$\ Ori   &  B0.5 Ia          &  39  &  3079 &   95 &  7.03  \\
HD 47240        &  B1 Ib            &  21  &  5452 &  100 & 32.42  \\
$o^2$\ CMa      &  B3 Ia            &  22  &  3001 &   50 & 85.51  \\
$\eta$\ CMa     &  B5 Ia            &  11  &  3030 &   60 & 44.65  \\
HD 64760        &  B0.5 Ib          & 208  &  6073 &  220 &  3.54  \\
$\rho$\ Leo     &  B1 Iab           &  12  &  6077 &   80 & 11.59  \\
HD 150168       &  B1 Ia            &  56  &  4944 &  120 &  6.13  \\
HD 164402       &  B0 Ib            &  62  &  5831 &   75 &  6.37  \\
HD 167756       &  B0.5 Ia          &  19  &  2811 &   79 &  8.66  \\ 
69 Cyg          &  B0 Ib            &  37  &  5151 &   75 & 13.81  \\ 
HD 219188       &  B0.5 II-III(n)   &  12  &  6024 &  150 &  3.28  \\  \hline
\end{tabular}
\end{center} 
\label{tab:stars}
\end{table}
\clearpage

\begin{table}[h]
\caption{{\bf Stellar properties}}
\begin{center}
\begin{tabular}{llrrrrr}\hline
Name      &  Sp Ty  &  \teff\ & $\log g$ & $v_{t}$ & $v_{rad}$  & \vinf  \\ \hline
$\xi$\ Per      &  O7.5III(n)((f))&  35000 &  3.5 &  10.0 &   65 & 2400   \\
HD 34656        &  O7 II(f)       &  35000 &  3.5 &  10.0 &  -12 & 2200   \\
HD 36486        &  O9.5 IINwk     &  30000 &  3.5 &   2.0 &   18 & 2050   \\
$\zeta$\ Pup    &  O4 I(n)fp      &  45000 &  3.0 &  10.0 &  -24 & 2500   \\
HD 93403        &  O5.5 III(fc)var&  40000 &  3.5 &  10.0 &  -15 & 2700   \\
HD 93843        &  O5 III(fc)     &  35000 &  3.5 &  10.0 &  -10 & 2700   \\
$\mu$\ Nor      &  O9.7 Iab       &  29000 &  3.5 &  10.0 &    6 & 2200   \\
HD 162978       &  O7.5 II((f))   &  35000 &  3.5 &  10.0 &  -12 & 2250   \\
HD 190429       &  O4 If+         &  39000 &  3.5 &  10.0 &  -16 & 2450   \\
68 Cyg          &  O7.5 III:n((f))&  36000 &  3.5 &  10.0 &    1 & 2450   \\
HD 207198       &  O9 Ib-II       &  32000 &  3.5 &  10.0 &  -18 & 2100   \\
19 Cep          &  O9.5 Ib        &  32000 &  3.5 &  10.0 &  -12 & 2050   \\ \hline
$\zeta$\ Per    &  B1 Ib          &  23000 &  3.5 &  10.0 &   21 & 1200   \\
$\epsilon$\ Ori &  B0 Ia          &  28000 &  3.0 &  10.0 &   27 & 1700   \\
$\kappa$\ Ori   &  B0.5 Ia        &  26000 &  3.5 &  10.0 &   20 & 1550   \\
HD 47240        &  B1 Ib          &  24000 &  3.0 &  10.0 &   33 & 1050   \\
$o^2$\ CMa      &  B3 Iab         &  17000 &  3.0 &  10.0 &   48 &  580   \\
$\eta$\ CMa     &  B5 Ia          &  16000 &  2.5 &  10.0 &   41 &  450   \\
HD 64760        &  B0.5 Ib        &  25000 &  2.5 &  10.0 &   41 & 1650   \\
$\rho$\ Leo     &  B1 Ib          &  25000 &  3.5 &  10.0 &   42 & 1300   \\
HD 150168       &  B1 Ia          &  25000 &  3.0 &  10.0 &    6 & 1385   \\
HD 164402       &  B0 Iab/b       &  28000 &  3.0 &  10.0 &    4 & 1680   \\
HD 167756       &  B0.5 Ib        &  28000 &  3.5 &  10.0 &  -25 & 2050   \\
69 Cyg          &  B0 Ib          &  27000 &  3.0 &  10.0 &    2 & 1800   \\
HD 219188       &  B0.5 III       &  28000 &  3.5 &   2.0 &   71 & 1500   \\
 \hline  
\end{tabular}
\end{center}
\label{tab:params}
\end{table}
\begin{table*}[h]
\caption{{\bf Series properties}}
\begin{center}
\begin{tabular}{lrrrrrrr}\hline
Name      &    N & $\Delta$ t (days) &  Date & $\tau$ & $\sigma(\tau)$ & $\sigma(\tau)/\tau$ & $\rho(\tau, r)$  \\ \hline
$\xi$ Per   &   33 &   3.59 &  9/87 & 1.22 & 0.17 & 0.14 &  -0.63 \\
$\xi$ Per   &   25 &   3.00 & 10/88 & 1.08 & 0.19 & 0.18 &  -0.66 \\
$\xi$ Per   &   36 &   4.42 & 10/91 & 1.21 & 0.24 & 0.20 &  -0.52 \\
$\xi$ Per   &   68 &   8.72 & 10/94 & 1.26 & 0.23 & 0.18 &  -0.39 \\
HD 34656    &   29 &   5.07 &  2/91 & 0.89 & 0.10 & 0.12 &   0.40 \\
$\zeta$ Pup &  139 &  15.86 &  1/95 & 1.28 & 0.10 & 0.08 &   0.54 \\
HD 93843    &   67 &  29.48 &  5/96 & 0.55 & 0.08 & 0.16 &  -0.34 \\
68 Cyg      &   29 &   3.39 &  9/87 & 1.05 & 0.14 & 0.13 &   0.13 \\
68 Cyg      &   23 &   2.56 & 10/88 & 1.05 & 0.15 & 0.15 &  -0.54 \\
68 Cyg      &   23 &   2.56 & 10/89 & 1.09 & 0.08 & 0.07 &   0.42 \\
68 Cyg      &   40 &   4.52 & 10/91 & 1.08 & 0.13 & 0.12 &   0.29 \\
68 Cyg      &   43 &   5.45 & 10/94a& 0.99 & 0.14 & 0.14 &   0.44 \\
68 Cyg      &   22 &   2.71 & 10/94b& 1.04 & 0.10 & 0.09 &   0.05 \\
19 Cep      &   18 &  10.60 & 10/94 & 1.45 & 0.18 & 0.12 &   0.43 \\
HD 64760    &   47 &   5.07 &  3/93 & 0.98 & 0.21 & 0.21 &  -0.52 \\
HD 64760    &  148 &  15.77 &  1/95 & 1.04 & 0.29 & 0.28 &  -0.61 \\
HD 164402   &   24 &   6.83 &  8/86 & 0.61 & 0.14 & 0.22 &  -0.52 \\
HD 164402   &   17 &  16.10 &  3/95 & 0.61 & 0.11 & 0.19 &  -0.76 \\
69 Cyg      &   15 &  16.00 &  10/93& 1.00 & 0.26 & 0.26 &  -0.83 \\ \hline
\end{tabular}
\end{center}
\label{tab:series}
\end{table*}




\begin{table*}[h]
\caption{{\bf Derived parameters}}
\begin{center}
\begin{tabular}{lcccccccr}\hline
Name       & $r$  & $\sigma(r)$ &  $w_D$ & $\sigma(w_D)$ & 
$\tau$ & $\sigma(\tau)$ & $\sigma(\tau)/\tau$ & $\rho(\tau,r)$ \\ \hline
$\xi$ Per  & 1.49 & 0.15 &  0.08 & 0.02 &  1.14 & 0.24 & 0.21 & -0.33\\ 
HD 34656       & 1.37 & 0.07 &  0.06 & 0.02 &  0.88 & 0.11 & 0.12 &  0.41\\
HD 36486       & 1.18 & 0.15 &  0.12 & 0.03 &  0.75 & 0.17 & 0.23 & -0.42\\  
$\zeta$ Pup    & 1.63 & 0.13 &  0.05 & 0.01 &  1.25 & 0.11 & 0.08 &  0.55\\  
HD 93403       & 1.23 & 0.21 &  0.10 & 0.01 &  1.12 & 0.16 & 0.14 &  0.72\\  
HD 93843       & 1.54 & 0.17 &  0.07 & 0.01 &  0.55 & 0.08 & 0.15 & -0.30\\  
$\mu$ Nor      & 1.19 & 0.13 &  0.13 & 0.02 &  1.71 & 0.24 & 0.14 &  0.54\\  
HD 162978      & 1.31 & 0.09 &  0.17 & 0.02 &  1.05 & 0.19 & 0.18 & -0.14\\  
HD 190429      & 1.12 & 0.08 &  0.09 & 0.01 &  0.99 & 0.08 & 0.08 &  0.71\\  
68 Cyg         & 1.46 & 0.09 &  0.07 & 0.02 &  1.04 & 0.13 & 0.12 &  0.12\\  
HD 207198      & 1.49 & 0.20 &  0.09 & 0.01 &  1.03 & 0.24 & 0.24 & -0.53\\  
19 Cep         & 1.32 & 0.09 &  0.14 & 0.02 &  1.38 & 0.22 & 0.16 & -0.29\\ \hline 
$\zeta$ Per    & 1.50 & 0.14 &  0.11 & 0.01 &  0.40 & 0.10 & 0.24 &  0.12\\  
$\epsilon$ Ori & 1.18 & 0.16 &  0.16 & 0.02 &  2.46 & 0.40 & 0.16 &  0.59\\  
$\kappa$ Ori   & 1.31 & 0.13 &  0.12 & 0.01 &  1.53 & 0.25 & 0.16 &  0.59\\  
HD 47240       & 1.21 & 0.19 &  0.11 & 0.02 &  2.12 & 0.40 & 0.19 & -0.21\\  
$o^2$ CMa      & 1.32 & 0.11 &  0.15 & 0.04 &  2.23 & 0.41 & 0.18 &  0.46\\  
$\eta$ CMa     & 1.64 & 0.28 &  0.13 & 0.03 &  1.27 & 0.58 & 0.45 & -0.64\\  
HD 64760       & 1.58 & 0.20 &  0.07 & 0.02 &  1.01 & 0.28 & 0.27 & -0.55\\  
$\rho$ Leo     & 1.46 & 0.11 &  0.13 & 0.01 &  1.53 & 0.49 & 0.32 & -0.20\\  
HD 150168      & 1.52 & 0.25 &  0.14 & 0.02 &  0.79 & 0.16 & 0.21 & -0.78\\  
HD 164402      & 1.42 & 0.18 &  0.16 & 0.01 &  0.60 & 0.12 & 0.19 & -0.34\\  
HD 167756      & 1.34 & 0.15 &  0.15 & 0.01 &  0.67 & 0.11 & 0.17 &  0.02\\  
69 Cyg         & 1.79 & 0.23 &  0.14 & 0.01 &  0.95 & 0.25 & 0.26 & -0.80\\  
HD 219188      & 1.29 & 0.23 &  0.05 & 0.02 &  0.69 & 0.24 & 0.35 & -0.61\\ \hline 
\end{tabular}
\end{center}
\label{tab:taus}
\end{table*}

\begin{table}[h]
\caption{{\bf Parameter Errors}}
\begin{center}
\begin{tabular}{lccccccc}\hline
Star         & SWP   & $\tau$ & $\sigma(\tau)/\tau $ & $r$  & $\sigma(r)/r$
& $w_D$  & $\sigma(w_D)/w_D$ \\ \hline
$\xi$~Per    & 52617 & 1.256 & 0.062 & 1.458 & 0.050 & 0.056 & 0.117 \\
HD 34656     & 40788 & 0.961 & 0.046 & 1.449 & 0.031 & 0.053 & 0.098 \\
HD 36486     & 28001 & 1.087 & 0.042 & 1.065 & 0.023 & 0.141 & 0.085 \\
$\zeta$~Pup  & 36143 & 1.121 & 0.057 & 1.587 & 0.051 & 0.043 & 0.160 \\
HD 93403     & 47534 & 1.138 & 0.060 & 1.436 & 0.039 & 0.105 & 0.101 \\
HD 93843     & 57062 & 0.620 & 0.044 & 1.675 & 0.055 & 0.099 & 0.082 \\
HD 149038    & 27939 & 2.251 & 0.071 & 1.266 & 0.050 & 0.130 & 0.031 \\
HD 162978    & 30502 & 1.140 & 0.057 & 1.339 & 0.039 & 0.172 & 0.047 \\
HD 190429    & 38970 & 0.971 & 0.065 & 1.051 & 0.051 & 0.059 & 0.094 \\
68 Cyg       & 42821 & 1.118 & 0.051 & 1.434 & 0.039 & 0.100 & 0.051 \\
HD 207198    & 26001 & 1.432 & 0.070 & 1.321 & 0.042 & 0.081 & 0.130 \\
19 cep       & 31769 & 1.371 & 0.043 & 1.365 & 0.030 & 0.144 & 0.051 \\
HD 24398     & 03053 & 0.389 & 0.041 & 1.601 & 0.068 & 0.057 & 0.230 \\
HD 37128     & 24879 & 2.082 & 0.065 & 1.243 & 0.037 & 0.126 & 0.068 \\
$\kappa$~Ori & 06733 & 1.273 & 0.046 & 1.250 & 0.028 & 0.106 & 0.046 \\
HD 47240     & 48873 & 2.214 & 0.065 & 1.530 & 0.040 & 0.089 & 0.040 \\
$o^2$ CMa    & 30153 & 2.434 & 0.076 & 1.328 & 0.064 & 0.174 & 0.075 \\
$\eta$ CMa   & 30198 & 2.544 & 0.094 & 1.287 & 0.076 & 0.159 & 0.083 \\
HD 64760     & 53742 & 1.443 & 0.049 & 1.378 & 0.032 & 0.083 & 0.031 \\
HD 91316     & 11312 & 1.751 & 0.052 & 1.565 & 0.029 & 0.137 & 0.033 \\
HD 150168    & 47222 & 0.842 & 0.057 & 1.511 & 0.048 & 0.151 & 0.032 \\
HD 164402    & 54132 & 0.647 & 0.047 & 1.495 & 0.069 & 0.156 & 0.082 \\
HD 167756    & 29025 & 0.687 & 0.055 & 1.364 & 0.058 & 0.152 & 0.020 \\
HD 204172    & 38979 & 0.820 & 0.043 & 1.922 & 0.040 & 0.131 & 0.059 \\
HD 219188    & 53057 & 1.219 & 0.073 & 1.259 & 0.026 & 0.069 & 0.071 \\
\hline
\end{tabular}
\end{center}
\label{tab:errors}
\end{table}

\end{document}